\documentclass[journal,onecolumn]{IEEEtran}
\usepackage{amsmath,amssymb,amsfonts}
\usepackage{amsthm}
\usepackage{mathtools}
\usepackage{graphicx}
\usepackage{array}
\usepackage{dsfont}
\usepackage{extarrows}
\usepackage[most]{tcolorbox}
\usepackage{xcolor}
\usepackage{cite}
\usepackage[hidelinks]{hyperref}

\ifCLASSINFOpdf

\else

\fi

\usepackage{fancyhdr}
\usepackage[calcwidth]{titlesec}
\usepackage[font={footnotesize},labelfont={bf}]{caption}
\usepackage{setspace}

\usepackage{extarrows}

\usepackage[most]{tcolorbox}
\newtcolorbox{myt}[2][]{%
  attach boxed title to top center
               = {yshift=-4pt},
  colback      = blue!5!white,
  colframe     = blue!75!black,
  halign       = flush left,
  fonttitle    = \bfseries\sffamily,
  colbacktitle = blue!65!black,
  title        = #2,#1,
  enhanced,
}
\newtcolorbox{myd}[2][]{%
  attach boxed title to top center
               = {yshift=-4pt},
  colback      = violet!5!white,
  colframe     = violet!75!black,
  halign       = flush left,
  fonttitle    = \bfseries\sffamily,
  colbacktitle = violet!65!black,
  title        = #2,#1,
  enhanced,
}
\newtcolorbox{mye}[2][]{%
  attach boxed title to top center
               = {yshift=-4pt},
  colback      = purple!5!white,
  colframe     = purple!75!black,
  halign       = flush left,
  fonttitle    = \bfseries\sffamily,
  colbacktitle = purple!65!black,
  title        = #2,#1,
  enhanced,
}

\newtcolorbox{myg}[2][]{%
  attach boxed title to top center
               = {yshift=-4pt},
  colback      = green!5!white,
  colframe     = green!75!black,
  halign       = flush left,
  fonttitle    = \bfseries\sffamily,
  colbacktitle = green!65!black,
  title        = #2,#1,
  enhanced,
}

\providecommand{\U}[1]{\protect\rule{.1in}{.1in}}

\usepackage{dsfont}

\usepackage{mathtools}

\def\1{\mathbf{1}}

\def\sub{{\rm sub}}

\def\prob{{\rm Prob}}

\def\cptp{{\rm CPTP}}

\def\ker{{\rm Ker}}

\def\upb{\underline{\mathbf{p}}}

\newcommand{\epm}{\end{pmatrix}}
\newcommand{\bpm}{\begin{pmatrix}}

\renewcommand{\log}{{\operatorname{log}}}

\newcommand{\ebm}{\end{bmatrix}}
\newcommand{\bbm}{\begin{bmatrix}}

\def\bmyd{\begin{myd}{}
\begin{definition}}
\def\emyd{\end{definition}\end{myd}}

\def\bmyl{\begin{myg}{}
\begin{lemma}}
\def\emyl{\end{lemma}\end{myg}}

\def\bmyt{\begin{myt}{}
\begin{theorem}}
\def\emyt{\end{theorem}\end{myt}}

\def\bmyc{\begin{myg}{}
\begin{corollary}}
\def\emyc{\end{corollary}\end{myg}}

\def\tmu{{\tilde{\mu}}}

\def\>{\rangle}
\def\<{\langle}

\def\mE{\mathcal{E}}

\newcommand{\supp}{\operatorname{supp}}

\renewcommand{\ge}{\geqslant}
\renewcommand{\le}{\leqslant}
\renewcommand{\geq}{\geqslant}
\renewcommand{\leq}{\leqslant}

\newcommand{\ben}{\begin{enumerate}}
\newcommand{\een}{\end{enumerate}}

\theoremstyle{definition}
\newtheorem{theorem}{Theorem}
\theoremstyle{definition}
\newtheorem{corollary}[theorem]{Corollary}
\theoremstyle{definition}
\newtheorem{lemma}[theorem]{Lemma}

\theoremstyle{definition}
\newtheorem{definition}{Definition}

\usepackage{array}
\usepackage{amsmath}

\newcommand{\bea}{\begin{eqnarray}}
\newcommand{\eea}{\end{eqnarray}}
\newcommand{\be}{\begin{equation}}
\newcommand{\ee}{\end{equation}}
\newcommand{\ba}{\begin{equation}\begin{aligned}}
\newcommand{\ea}{\end{aligned}\end{equation}}

\newcommand{\bee}{\begin{enumerate}}
\newcommand{\eee}{\end{enumerate}}

\def\be{\begin{equation}}
\def\ee{\end{equation}}

\newcommand{\mb}{\mathfrak{B}}
\newcommand{\md}{\mathfrak{D}}

\newcommand{\D}{\mathbf{D}}
\newcommand{\Ediv}{\mathbf{E}}

\newcommand{\lr}{\rangle\langle}

\newcommand{\tr}{{\rm Tr}}

\newcommand{\da}{\downarrow}

\newcommand{\eps}{\varepsilon}

\newcommand{\mbb}[1]{\mathbb{#1}}

\newcommand{\eqdef}{\coloneqq}

\def\r{\mathbf{r}}
\def\s{\mathbf{s}}
\def\p{\mathbf{p}}
\def\q{\mathbf{q}}

\def\e{\mathbf{e}}

\def\t{\mathbf{t}}
\def\u{\mathbf{u}}

\def\0{\mathbf{0}}

\def\tD{\widetilde{D}}

\def\tQ{\widetilde{Q}}

\def\trho{{\tilde{\rho}}}

\begin{document}

% Title and authors
\title{Optimal Universal Bounds for Quantum Divergences}

\author{Gilad~Gour%
\thanks{G. Gour is with the Faculty of Mathematics, Technion -- Israel Institute of Technology, Haifa 3200003, Israel. Email: giladgour@gmail.com.}%
}

\maketitle

\begin{abstract}
We identify a universal structural principle underlying the smoothing of classical divergences: the optimizer of the smoothing problem is a clipped probability vector, independently of the specific divergence. This yields a divergence-independent characterization of all smoothed classical divergences and reveals a common geometric structure behind seemingly different quantities. Building on this structural insight, we derive optimal universal bounds for smoothed quantum divergences, including quantum R\'enyi divergences of arbitrary order and the hypothesis-testing divergence. Our inequalities relate divergences of different orders through bounds of the form $D_\beta^{\varepsilon} \le D_\alpha + \mathrm{correction}$ and $D_\beta^{\varepsilon} \ge D_\alpha + \mathrm{correction}$, and we prove that the correction terms are optimal among all universal, state-independent inequalities of this type. Consequently, our results strictly improve previously known bounds whenever those were suboptimal, and in cases where earlier bounds coincide with ours, our analysis establishes their optimality. In particular, we obtain optimal universal bounds for the hypothesis-testing divergence.
\end{abstract}


%\tableofcontents

\section{Introduction}

Universal bounds are inequalities that relate smoothed entropic or divergence quantities to reference information measures, with correction terms independent of the states and dimension; see, for example,~\cite{Renner2005,Hayashi2006,Tomamichel2015,KW2024,Gour2025}. Additive reference measures are especially useful for asymptotic applications, but additivity is not part of the definition of a universal bound. In particular, measured divergences can serve as reference quantities even when they are not additive.
The hallmark of a universal bound is that it is independent of both the Hilbert space dimension and the particular states under consideration. Rather than exploiting special structural features of a given system, it provides an inequality of the form
\be
D_\beta^\varepsilon(\rho\|\sigma)
\;\lessgtr\;
D_\alpha(\rho\|\sigma)
+ \mathrm{correction}(\varepsilon,\alpha,\beta),
\ee
where the correction term depends only on the smoothing parameter and the divergence orders, and not on the dimension or on the states themselves. Such relations provide uniform control of operationally relevant quantities in the single-shot regime.

This perspective provides a direct conceptual bridge between the single-shot and asymptotic regimes. For example, universal lower bounds on the hypothesis-testing divergence immediately yield a streamlined proof of the quantum Stein's lemma~\cite{WR2012,ON2000,OH2001Stein}: combining dimension-independent one-shot inequalities with additivity under tensor powers shows that the optimal asymptotic error exponent is governed by the relative entropy. In this sense, universal bounds isolate the structural mechanism by which asymptotic distinguishability emerges from single-shot considerations.

Smooth divergences and the relations between them have been studied extensively; see, for example,~\cite{Tomamichel2015,Datta2009,WR2012} and references therein. In particular, a series of works has developed refined inequalities connecting the hypothesis-testing divergence, the smooth max-relative entropy, and various R\'enyi divergences. Recent progress in this direction was reported in~\cite{RLD2025}. In particular, Section~VIII\,A of~\cite{RLD2025} establishes strengthened bounds relating the smooth max-relative entropy to R\'enyi divergences, which in turn lead to improved constraints on the hypothesis-testing divergence.

However, two fundamental questions remain open. First, are such universal inequalities optimal? In other words, is the correction term minimal among all bounds of the same functional form that are independent of both the dimension and the underlying states? Second, can one obtain universal control over the smoothing of R\'enyi divergences of \emph{arbitrary order}, rather than restricting attention to particular cases such as the max-relative entropy $D_{\max}$ or the hypothesis-testing divergence $D_H$?

The latter issue is particularly important for applications. Several central tools in quantum information theory, most notably the decoupling theorem~\cite{Dupuis2010} and the convex-split lemma~\cite{ADJ2017,Gour2025b}, are naturally governed by the collision relative entropy, i.e., the R\'enyi divergence of order $2$. In practice, however, smoothing arguments often proceed only indirectly: one first bounds the relevant R\'enyi divergence in terms of $D_{\max}$ or the hypothesis-testing divergence, performs smoothing at that level, and then translates the resulting bound back to the R\'enyi quantity of interest. This detour arises because direct universal control over smoothed R\'enyi divergences has so far been unavailable.

Our results provide precisely this missing ingredient. We establish universal bounds with additive correction terms that apply directly to smoothed R\'enyi divergences of arbitrary order. This makes it possible to smooth quantities such as the collision divergence in a dimension-independent and optimal manner, without passing through intermediate divergences. As a result, applications such as decoupling and convex-split can be analyzed directly at the R\'enyi parameter that naturally governs the task.

For the comparisons studied here, we also resolve the optimality problem. We prove that the correction terms are minimal among all state- and dimension-independent bounds with the same divergence quantities and functional form. Within this class, our results improve earlier nonoptimal correction terms and establish optimality whenever the earlier terms coincide with ours.\\

\noindent To illustrate the scope of our results, we highlight the following three examples:

\begin{itemize}
\item \textbf{Example 1:}
Optimal lower bounds for the hypothesis-testing divergence.
A universal lower bound for the hypothesis-testing divergence was obtained in~\cite{AMV2012} (see also~\cite{FGW2025}). For every $\eps,\alpha\in(0,1)$ it states that
\be\label{oub0}
D_{H}^\eps(\rho\|\sigma)\geq D_{\alpha}(\rho\|\sigma)+\log\frac1{1-\alpha}-\frac{\alpha}{1-\alpha}\log\left(\frac\alpha\eps\right)\;.
\ee
A related universal lower bound was derived in~\cite{RLD2025}, with a correction term that is not uniformly comparable to the one in~\eqref{oub0}. 
Here we show that for $\eps<\alpha$ the bound in~\eqref{oub0} is in fact optimal.
More precisely, we prove that
\begin{align}\label{oub1}
&D_{H}^\eps(\rho\|\sigma)
\geq\\ 
&D_{\alpha}(\rho\|\sigma)+
\begin{cases}
\log\frac1{1-\eps} & \text{if }\alpha\in[0,\eps]\\
\log\frac1{1-\alpha}-\frac{\alpha}{1-\alpha}\log\left(\frac\alpha\eps\right) & \text{if }\alpha\in(\eps,1)
\end{cases}\nonumber
\end{align}
is the optimal universal lower bound.
Consequently, whenever the bounds of~\cite{RLD2025} and~\cite{AMV2012} differ from~\eqref{oub1}, they are strictly weaker. 
Interestingly, we show that optimality continues to hold even if the Petz R\'enyi divergence $D_\alpha$ is replaced by the measured R\'enyi divergence $D_\alpha^{\mathbb{M}}$, which is known to be the minimal quantum extension~\cite{GT2020,Matsumoto2018}.
\item \textbf{Example 2:} Optimal bounds for smoothed max-relative entropy.
Corollary~14 of~\cite{RLD2025} established the universal bound
\be\label{rld2}
\widetilde{D}_{\max}^\eps(\rho\|\sigma)
\leq
D_{\alpha}^{\mathbb{M}}(\rho\|\sigma)
+\frac{1}{\alpha-1}\log\frac1\eps-\log\frac1{1-\eps}\;,
\ee
for the modified smoothed max-relative entropy $\widetilde{D}_{\max}^\eps$ defined via the information-spectrum divergence.
Our results determine the optimal universal upper bound, which is given by
\begin{align}
&\widetilde{D}_{\max}^\eps(\rho\|\sigma)\leq\\
&
{D_{\alpha}^{\mathbb{M}}(\rho\|\sigma)+
\begin{cases}
{\log}(1-\eps) & \text{if }\eps\geq\frac1\alpha\\
\frac1{\alpha-1}\log\frac1\eps
-\frac{\alpha\log\alpha}{\alpha-1}-\log\frac1{\alpha-1}
& \text{if }\eps<\frac1\alpha
\end{cases}}\nonumber
\end{align}
One can verify that this correction term is strictly smaller than the one in~\eqref{rld2}, indicating that the bound of~\cite{RLD2025} is not optimal.
\item \textbf{Example 3:} Bounds for the smoothed collision divergence.
As a further application, we obtain universal bounds for smoothed R\'enyi divergences of arbitrary order. 
In particular, for $1<\alpha\leq2$ we prove 
\begin{align}\label{mainforsub2}
&D_2^{\eps,\le,\mathbb{M}}(\rho\|\sigma)
\leq D_\alpha^{\mathbb{M}}(\rho\|\sigma)\\
&+\begin{cases}
\frac{2-\alpha}{\alpha-1}\log\left(\frac{2-\alpha}{\alpha\eps}\right)
+
2\log\left(\frac{2(\alpha-1)}\alpha\right)&\text{if }\;\;\eps\le\frac{2-\alpha}{\alpha}\\
2\log(1-\eps)&\text{if }\;\;\eps>\frac{2-\alpha}{\alpha}
\end{cases}
\,.\nonumber
\end{align}
where the superscript $\le$ stands for smoothing with subnormalized states. The above bound is uniformly tighter than the estimate in Lemma~10 of~\cite{RT2026}. A subsequent version of that work also provides an alternative direct proof of the first-branch collision bound.
Since the collision divergence naturally governs tools such as the quantum decoupling theorem and the convex-split lemma, this bound can be used to sharpen existing performance costs in these protocols~\cite{RT2026}.
\end{itemize}

The clipped structure that emerges from the smoothing optimization is closely related
to thresholding phenomena that appear in other convex optimization problems.
In particular, separable convex objectives under $\ell_1$-type constraints are known
to admit ``clipping'' or thresholding solutions, a structure that arises, for
example, in Euclidean projections onto the probability simplex and weighted
$\ell_1$-balls~\cite{Duchi2008,Condat2016}.
Related extremal phenomena also appear in information theory in the study of
$f$-divergence inequalities under prescribed total variation constraints, where
optimization over pairs of distributions often leads to extremal binary
solutions~\cite{Gilardoni2010,Sason2016}.
However, the smoothing problem studied here is of a fundamentally different
nature: rather than optimizing jointly over distributions, we fix both the
reference distribution $\q$ and the center $\p$, and minimize the divergence
over all distributions $\p'$ lying within a total-variation ball around $\p$.
To the best of our knowledge, an explicit characterization of the optimal
distribution in this fixed-center smoothing problem has not previously
appeared in the literature.

A key structural ingredient underlying our approach is a complete solution of the classical smoothing problem. Our analysis relies heavily on tools from majorization theory and its relative variant~\cite{MOA2011,Gour2025}. Concretely, the framework rests on three main ingredients:

\begin{itemize}
\item[(i)] \emph{Flattest approximations via relative majorization.}
Building on flattest approximations under majorization~\cite{HY2010,HD2021,HOS2018}, we construct a clipped probability vector that is minimal under relative majorization throughout the total-variation ball. It therefore minimizes every classical divergence, giving a divergence-independent solution of the classical smoothing problem.

\item[(ii)] \emph{Classical reductions and binary extremizers.}
Relative majorization allows us to replace a positive rational reference distribution by a uniform one in a larger dimension; an approximation argument extends this reduction to the universal constants studied here. For Theorems~\ref{thmg1} and~\ref{thmg1sub}, we instead use a split--merge construction: splitting intermediate likelihood ratios and then merging equal-ratio coordinates produces a binary pair whose objective is at least as large. Thus binary distributions suffice to approach the optimal corrections in these theorems.

\item[(iii)] \emph{Classical-quantum reduction through measured divergences.}
On the quantum side, we systematically reduce smoothing problems to classical ones via measured divergences. By analysing the measured versions of the relevant quantum divergences and exploiting their exact classical structure, we are able to lift optimal classical bounds to the fully quantum setting. In this way, the classical clipped-vector solution becomes the backbone of our optimal universal bounds for quantum divergences.
\end{itemize}

An important structural aspect of our approach concerns the choice of smoothing metric. 
Throughout the classical analysis we employ smoothing with respect to the trace distance 
(total variation distance in the classical setting). This choice is not merely technical. 
The trace-distance $\varepsilon$-ball around a probability vector admits maximal and minimal 
elements under majorization, which are precisely the steepest and flattest (clipped) 
approximations~\cite{HY2010,HD2021,HOS2018}. In other words, the trace-distance ball is compatible with the order structure 
induced by majorization.

This order-theoretic compatibility is essential. 
It is precisely the existence of extremal elements in the $\varepsilon$-ball 
that allows the smoothing optimization to admit a universal clipped-vector 
solution. By contrast, smoothing defined via the purified distance or fidelity 
does not preserve this order-theoretic structure: the corresponding 
$\varepsilon$-balls need not admit extremal elements under majorization, and 
therefore do not lead to a comparable universal characterization. In this sense, 
the trace distance is the natural metric for analysing smoothing through the 
lens of majorization theory.

The remainder of the paper is organized as follows. 
In Section~\ref{Sec2} we state the main results, including optimal universal upper and lower bounds for smoothed R\'enyi divergences and the hypothesis-testing divergence, as well as a closed-form characterization of smoothed classical divergences. 
Section~\ref{Sec3} reviews the necessary preliminaries on majorization, classical and quantum divergences, and smoothing. 
In Section~\ref{Sec4} we develop the structural framework underlying our results, showing how smoothing problems can be reduced to classical optimization problems through relative majorization and measurement reductions. 
Sections~\ref{Sec5}–\ref{Sec9} contain the proofs of the main theorems, including the optimal universal bounds and the explicit characterization of smoothed classical divergences.

\section{Main Results}\label{Sec2}

In this section we present the main results of the paper. We determine the optimal universal bounds relating smoothed R\'enyi divergences to unsmoothed divergences of different orders and explicitly compute the corresponding correction terms. These bounds are shown to be optimal among all dimension-independent and state-independent inequalities of the same form. We also derive optimal bounds for the hypothesis-testing divergence and establish a closed-form expression for smoothed classical divergences. Precise notation and definitions will be introduced in the next section.
Whenever an optimal universal constant is written as a supremum of a difference of divergences, the supremum is taken over pairs for which the subtracted divergence is finite. Pairs where it is infinite impose no restriction on the corresponding upper-bound inequality. Thus no expression of the form $\infty-\infty$ is used.

\subsection{Optimal Universal Upper Bounds of Smoothed R\'enyi Divergences}

Let $\eps\in(0,1)$ and $\alpha,\beta\in[0,\infty]$. 
Let $\D_\alpha$ be any quantum extension of the classical R\'enyi divergence of order $\alpha$, and let
$\D = D_{\beta}^{\eps,\mathbb{M}}$ denote the minimal (also known as \emph{measured}) quantum extension of the smoothed classical R\'enyi divergence $D_\beta^\eps$.
Define
\be\label{mainex}
\mu(\eps,\alpha,\beta)\eqdef\sup_A\sup_{\rho,\sigma\in\md(A)}\Big\{D_\beta^{\eps,\mathbb{M}}(\rho\|\sigma)- \D_\alpha(\rho\|\sigma)\Big\}\;.
\ee
By definition, $\mu(\eps,\alpha,\beta)$ is the smallest number satisfying
\be\label{uni8}
D_\beta^{\eps,\mathbb{M}}(\rho\|\sigma)\leq \D_\alpha(\rho\|\sigma)+\mu(\eps,\alpha,\beta)\qquad\forall\;\rho,\sigma\in\md(A)\;.
\ee
In the theorem below we make use of the parameter $\theta\in(0,1]$ defined by
\be\label{theta}
\theta\eqdef
\begin{cases}
\frac{\beta-\alpha}{\beta(1-\alpha)} &\text{if }0\le\alpha<\beta<1\\
\frac{\beta-\alpha}{\alpha(\beta-1)} &\text{if }\beta>\alpha>1
\end{cases}\;.
\ee
Let $\eps\in(0,1)$, $\alpha,\beta\in[0,\infty]$, and $h_2(\theta)\eqdef-\theta\log(\theta)-(1-\theta)\log(1-\theta)$ be the binary Shannon entropy of $\theta$. Then:
\bmyt\label{thmg1}
\be\label{mainfor}
\mu(\eps,\alpha,\beta)=\max\{0,\tmu\}
\ee
where
\be
\tmu\eqdef\begin{cases}
\frac{\beta}{1-\beta}\left(\theta\log\left(\frac1\eps\right)-h_2(\theta)\right)&\text{if }0\le\alpha<\beta<1\\
\frac{\alpha}{\alpha-1}\left(\theta\log\left(\frac1\eps\right)-h_2(\theta)\right)&\text{if }\beta>\alpha>1\\
0&\text{if }\alpha\geq \beta\geq 0\\
\infty & \text{otherwise}
\end{cases}
\ee
\emyt

Here $0\log 0=0$; in particular, $h_2(1)=0$ when $\alpha=0$. The expression above therefore gives the optimal correction term appearing in the universal bound~\eqref{uni8}.

For $\beta=\infty$ we have $D_\beta^{\eps,\mathbb{M}}=D_{\max}^{\eps,\mathbb{M}}$ and $\theta=1/\alpha$. 
In this case Theorem~\ref{thmg1} yields, for every $\rho,\sigma\in\md(A)$ and $\alpha>1$,
\ba\label{wild}
&D_{\max}^{\eps,\mathbb{M}}(\rho\|\sigma)
\leq 
\D_{\alpha}(\rho\|\sigma)\\
&\quad+
\max\left\{
0\;,\;
\frac1{\alpha-1}\log\frac1\eps
-\frac{\alpha\log(\alpha)}{\alpha-1}
-\log\frac1{\alpha-1}
\right\}.
\ea
In particular, the correction term satisfies $\mu(\eps,\alpha,\infty)>0$ whenever 
$0<\eps<(\alpha-1)^{\alpha-1}/\alpha^\alpha$, while $\mu(\eps,\alpha,\infty)=0$ otherwise.

\subsection{Optimal Universal Lower Bounds for Smoothed R\'enyi Divergences}

Let $\eps\in(0,1)$ and $\alpha,\beta\in[0,\infty]$. 
Let $\D_\beta$ be any quantum extension of the classical R\'enyi divergence of order $\beta$, let $\D_\beta^\eps$ denote its smoothed version (see~\eqref{smoothed}), and let $D_{\alpha}^{\mathbb{M}}$ be the minimal (measured) quantum extension of the classical R\'enyi divergence $D_\alpha$. 
Define
\be
\nu(\eps,\alpha,\beta)
\eqdef
\sup_A\sup_{\rho,\sigma\in\md(A)}
\left\{
D_\alpha^{\mbb{M}}(\rho\|\sigma)-\D_\beta^\eps(\rho\|\sigma)
\right\}.
\ee
By definition, $\nu(\eps,\alpha,\beta)$ is the smallest dimension-independent number such that
\be
\D_\beta^\eps(\rho\|\sigma)
\ge
D_\alpha^{\mbb{M}}(\rho\|\sigma)-\nu(\eps,\alpha,\beta)
\qquad
\forall\;\rho,\sigma\in\md(A).
\ee
Then:
\bmyt\label{thmg2}
\be
\nu(\eps,\alpha,\beta)=
\begin{cases}
\left(\frac1{\beta-1}+\frac1{1-\alpha}\right)\log\frac1{1-\eps}
& \text{if }\beta>1>\alpha\\
\infty & \text{otherwise}
\end{cases}
\ee
\emyt

The measured comparator $D_\alpha^{\mathbb{M}}$ need not be additive. The bound is universal because its correction term $\nu(\eps,\alpha,\beta)$ is independent of the states and dimension. Additive quantities can enter an asymptotic application through regularization, as explained in Section~III.C. The connection with strong converse exponents, and the role of the smoothing convention, are discussed after the proof of Theorem~\ref{thmg2} in Section~VI.

\subsection{Optimal Universal Bounds for the Hypothesis-Testing Divergence}

The optimal universal upper bound for the hypothesis-testing divergence is defined through the function
\be
\mu_H(\eps,\alpha)\eqdef
\sup_A\sup_{\rho,\sigma\in\md(A)}
\left\{
D_H^{\eps}(\rho\|\sigma)- \D_\alpha(\rho\|\sigma)
\right\},
\ee
where $\D_\alpha$ is any quantum extension of the R\'enyi relative entropy of order $\alpha\in[0,\infty]$.
For $\alpha>1$, as discussed in the introduction, it is known that for every pair of states $\rho,\sigma\in\md(A)$,
\be\label{alpha}
D_H^{\eps}(\rho\|\sigma)
\le
\D_\alpha(\rho\|\sigma)
+
\frac\alpha{\alpha-1}\log\!\left(\frac1{1-\eps}\right).
\ee
Consequently,
\be
\mu_H(\eps,\alpha)
\le
\frac\alpha{\alpha-1}\log\!\left(\frac1{1-\eps}\right).
\ee
The following theorem shows that equality holds, and hence the universal bound~\eqref{alpha} is optimal.

\bmyt\label{thmg3}
For all $\eps\in(0,1)$ and $\alpha\in[0,\infty]$,
\be
\mu_H(\eps,\alpha)=
\begin{cases}
\frac\alpha{\alpha-1}\log\!\left(\frac1{1-\eps}\right) & \text{if }\alpha>1,\\
\infty & \text{otherwise}.
\end{cases}
\ee
\emyt

For the hypothesis-testing divergence there is also a well-known lower bound valid for all $\alpha\in(0,1)$~\cite{AMV2012} (see also~\cite{FGW2025} for an alternative derivation):
\be\label{oub}
D_{H}^\eps(\rho\|\sigma)
\ge
D_{\alpha}(\rho\|\sigma)
+
\log\frac1{1-\alpha}
-
\frac{\alpha}{1-\alpha}\log\!\left(\frac\alpha\eps\right).
\ee

In this paper we show that this lower bound is optimal for $\alpha\ge\eps$, even if the Petz R\'enyi divergence is replaced by the minimal quantum extension $D_\alpha^{\mbb{M}}$.
To this end we define
\be\label{200}
\nu_H(\eps,\alpha)
\eqdef
\sup_A\sup_{\rho,\sigma\in\md(A)}
\left\{
D_\alpha^{\mbb{M}}(\rho\|\sigma)-D^{\eps}_H(\rho\|\sigma)
\right\}
\;.
\ee
By definition, $\nu_H(\eps,\alpha)$ is the smallest dimension-independent number satisfying
\be
D^\eps_H(\rho\|\sigma)
\ge
D_\alpha^{\mbb{M}}(\rho\|\sigma)-\nu_H(\eps,\alpha)
\qquad
\forall\;\rho,\sigma\in\md(A).
\ee

\bmyt\label{thmg4}
For all $\eps\in(0,1)$,
\be
\nu_H(\eps,\alpha)=
\begin{cases}
-\log\frac1{1-\eps}
& \text{if }\alpha\in[0,\eps],\\
\frac{\alpha}{1-\alpha}\log\!\left(\frac\alpha\eps\right)-\log\frac1{1-\alpha}
& \text{if }\alpha\in(\eps,1),\\
\infty
& \text{if }\alpha\in[1,\infty].
\end{cases}
\ee
\emyt

\subsection{Optimal Universal Bounds with Smoothing over Subnormalized States}\label{sec:subnormalized}

Smoothing over subnormalized states provides a natural extension of the standard smoothing paradigm in quantum information theory. Allowing subnormalized states incorporates a controlled probability of failure, which arises naturally in one-shot tasks such as hypothesis testing and state discrimination. This relaxation enlarges the feasible set while preserving convexity, and often yields tighter and more tractable bounds. It also aligns well with variational characterizations of divergences and frequently admits explicit optimizers, making it particularly convenient for finite-size analysis.

We follow~\cite{RLD2025,RT2026} and use the generalized trace distance defined for two subnormalized states $\rho$ and $\sigma$ as
\be
\|\rho-\sigma\|_+\eqdef\frac12\big|\tr[\rho-\sigma]\big|+\frac12\|\rho-\sigma\|_1.
\ee
Here $(\rho-\sigma)_+$ denotes the positive part of $\rho-\sigma$. If $\tr\rho\ge\tr\sigma$, then $\|\rho-\sigma\|_+=\tr(\rho-\sigma)_+$.

We use the following convention for R\'enyi divergences on nonzero subnormalized operators. For $R=\gamma\rho$ and $S=\delta\sigma$, where $\rho,\sigma$ are normalized and $\gamma,\delta>0$, and for $\alpha\in(0,1)\cup(1,\infty)$, set
\be\label{subnormalized-homogeneity}
\D_\alpha(R\|S)
\eqdef\D_\alpha(\rho\|\sigma)
+\frac{\alpha}{\alpha-1}\log\gamma-\log\delta.
\ee
Thus the classical power sum, the Petz trace formula, and the usual sandwiched trace formula are used without division by $\tr R$. The measured and regularized measured extensions use the same convention. At the endpoints we set
\be
D_0(R\|S)=-\log\tr(\Pi_R S),\qquad
D_{\max}(R\|S)=\inf\{\log t:R\le tS\},
\ee
where $\Pi_R$ is the support projection of $R$. In particular, the first-argument scaling coefficient at order $\infty$ is one. No order-one subnormalized divergence is needed in Theorem~\ref{thmg1sub}; the convention~\eqref{subnormalized-homogeneity} does not have a finite order-one limit when $\gamma<1$. All feasible first arguments below are nonzero because their trace is at least $1-\eps>0$. Negative values are allowed under this convention.

For two \emph{normalized} density matrices $\rho$ and $\sigma$ and a divergence $\D$ with the specified subnormalized extension, we define
\be
\D^{\eps,\le}(\rho\|\sigma)\eqdef\inf\big\{\D(\trho\|\sigma)\;:\;\trho\geq 0\;,\quad\tr[\trho]\leq 1\;,\quad \|\rho-\trho\|_+\leq\eps\big\}\;.
\ee

Now, let $D_\beta^{\eps,\le}$ be the classical smoothed R\'enyi divergence, with $D_\beta^{\eps,\le,\mathbb{M}}$ its minimal (i.e., measured) quantum extension. Define
\be\label{mainexsub}
\mu_{\sub}(\eps,\alpha,\beta)\eqdef\sup_A\sup_{\rho,\sigma\in\md(A)}\Big\{D_\beta^{\eps,\le,\mathbb{M}}(\rho\|\sigma)- \D_\alpha(\rho\|\sigma)\Big\}\;.
\ee
By definition, $\mu_\sub(\eps,\alpha,\beta)$ is the smallest number satisfying
\be\label{uni8sub}
D_\beta^{\eps,\le,\mathbb{M}}(\rho\|\sigma)\leq \D_\alpha(\rho\|\sigma)+\mu_\sub(\eps,\alpha,\beta)\qquad\forall\;\rho,\sigma\in\md(A)\;.
\ee

In the theorem below we use the same parameter $\theta\in(0,1)$ defined in~\eqref{theta}. Let $\eps\in(0,1)$, $\alpha,\beta\in[0,\infty]$, and $h_2(\theta)\eqdef-\theta\log(\theta)-(1-\theta)\log(1-\theta)$ denote the binary Shannon entropy. Then:
\bmyt\label{thmg1sub}$\;$\\
\noindent\textbf{1.} For $\beta>\alpha>1$
\be\label{mainforsub}
\mu_\sub(\eps,\alpha,\beta)=\begin{cases}
\frac{\alpha}{\alpha-1}\left(\theta\log\left(\frac1\eps\right)-h_2(\theta)\right)&\text{if }\;\;\eps\le\theta\\
\frac\beta{\beta-1}\log(1-\eps)&\text{if }\;\;\eps>\theta
\end{cases}
\ee
\noindent\textbf{2.} For $0<\alpha<\beta<1$, $\mu_\sub(\eps,\alpha,\beta)=\mu(\eps,\alpha,\beta)$, where $\mu(\eps,\alpha,\beta)$ is given in~\eqref{mainfor}.
\emyt

For $\beta>\alpha>1$, the subnormalized correction agrees with the normalized correction whenever the first branch of~\eqref{mainforsub} is nonnegative. When that branch is negative, subnormalized smoothing already gives a strictly smaller correction; for $\eps>\theta$ the second branch applies.

In~\eqref{wild} we gave the optimal universal upper bound for $D_{\max}^{\eps,\mathbb{M}}$.
As an illustration of the theorem, consider the minimal extension $D_{\max}^{\eps,\le,\mathbb{M}}$, corresponding to $\beta=\infty$, which yields the divergence $\widetilde{D}_{\max}^\eps$. In this case $\theta=1/\alpha$. Using the shorthand $\mu_\sub(\eps,\alpha,\infty)\eqdef\mu_\sub(\eps,\alpha)$, we obtain that $\mu_{\sub}(\eps,\alpha)$ is the smallest number satisfying
\be\label{uni3}
\widetilde D^\eps_{\max}(\rho\|\sigma)
\leq
\D_{\alpha}(\rho\|\sigma)+\mu_{\sub}(\eps,\alpha)
\qquad
\forall\;\rho,\sigma\in\md(A)\;.
\ee
Taking the limit $\beta\to\infty$ in~\eqref{mainforsub} gives
\be\label{mueps} \mu_{\sub}(\eps,\alpha)= \begin{cases}  \frac1{\alpha-1}\log\frac1\eps -\frac{\alpha\log(\alpha)}{\alpha-1} -\log\frac1{\alpha-1} & \text{if }\eps<\frac1\alpha\\
\log(1-\eps) & \text{if }\eps\geq\frac1\alpha \end{cases} 
\ee
Another notable example is $\beta=2$ with $1<\alpha<2$. In this case $\theta=(2-\alpha)/\alpha$, recovering the bound in~\eqref{mainforsub2}.

\subsection{Closed Formula for Smoothed Classical Divergences}

Finally, we show that the classical smoothing problem admits
a universal optimizer, independently of the divergence.
Let $\eps\in[0,1]$ and $\p,\q\in\prob(d)$.
Denote by $\p^{(\eps)}$ the clipped probability vector
defined in Sec.~\ref{sec:arm}, with components given
in~\eqref{pstar} in the nontrivial regime
$0<\eps<\frac12\|\p-\q\|_1$.
For $\eps=0$ we set $\p^{(0)}=\p$, while
$\p^{(\eps)}=\q$ whenever
$\eps\ge\frac12\|\p-\q\|_1$.
The definition also covers reference distributions $\q$
with zero components.
Then the following holds.

\bmyt\label{thmg0}
Every classical divergence $\D$ satisfies
\be
\D^\eps(\p\|\q)
=
\D\big(\p^{(\eps)}\big\|\q\big),
\ee
where $\D^\eps$ denotes the $\eps$-smoothed variant
of $\D$ defined in~\eqref{csmoothed}.
\emyt

This result follows from the fact that $\p^{(\eps)}$
is the flattest $\eps$-approximation of $\p$ relative
to $\q$. Specifically,
\be
(\p',\q)\succ(\p^{(\eps)},\q)
\qquad
\forall\,\p'\in\mb^\eps(\p).
\ee
Consequently, DPI implies that the same vector
$\p^{(\eps)}$ minimizes every classical divergence
over $\mb^\eps(\p)$.
The construction is given in Sec.~\ref{sec:arm},
and the proof appears in Sec.~\ref{Sec9}.

\section{Preliminaries}\label{Sec3}

\subsection{Majorization and Relative Majorization}\label{secmajo}

Let $d\in\mbb{N}$ and $\prob(d)$ be the set of all $d$-dimensional probability vectors.  
For $\p\in\prob(d)$ we denote by $\p^\da$ the vector obtained by rearranging the components of $\p$ in non-increasing order, and by $\prob^\da(d)$ the set of vectors satisfying $\p=\p^\da$.  
For $\p,\q\in\prob(d)$ we say that $\p$ majorizes $\q$, written $\p\succ\q$, if
\be
\|\p\|_{(k)}\ge \|\q\|_{(k)}\;,\qquad \forall\,k\in[d],
\ee
where $[d]\eqdef\{1,\ldots,d\}$ and
\be
\|\p\|_{(k)}\eqdef\sum_{x\in[k]} p_x^\da\;,
\ee
is the Ky–Fan norm. We denote by $\u\eqdef(1/d,\ldots,1/d)^T$ the uniform distribution in $\prob(d)$ and by $\e_1\eqdef(1,0,\ldots,0)^T$ a point mass.

For $\eps\in[0,1]$ and $\p\in\prob(d)$ define the $\eps$-ball
\be
\mb^\eps(\p)\eqdef\left\{\p'\in\prob(d):\frac12\|\p-\p'\|_1\le\eps\right\}.
\ee
Remarkably, $\mb^\eps(\p)$ has minimal and maximal elements under majorization, known as the \emph{flattest} and \emph{steepest} $\eps$-approximations of $\p$~\cite{HY2010,HD2021,HOS2018} (see also~\cite[Ch.~4]{Gour2025}).

The \emph{flattest} $\eps$-approximation $\p^{(\eps)}$ of $\p$ is the vector in $\mb^\eps(\p)$ satisfying $\p^{(\eps)}\prec\p'$ for all $\p'\in\mb^\eps(\p)$.  
Since $\u\prec\q$ for all $\q\in\prob(d)$, if $\u\in\mb^\eps(\p)$ then $\p^{(\eps)}=\u$.  
We therefore assume
\be\label{pue}
\frac12\|\p-\u\|_1>\eps .
\ee
Let $\p\in\prob^\da(d)$ and define
\be\label{a}
a\eqdef\max_{\ell\in[d]}\frac{\|\p\|_{(\ell)}-\eps}{\ell}
=\frac{\|\p\|_{(k)}-\eps}{k},
\ee
where $k$ is the largest index achieving the maximum. Similarly,
\be\label{b}
b\eqdef\min_{\ell\in[d-1]}\frac{1-\|\p\|_{(\ell)}+\eps}{d-\ell}
=\frac{1-\|\p\|_{(m)}+\eps}{d-m},
\ee
where $m$ is the smallest index achieving the minimum.  
If~\eqref{pue} holds then $k\le m$ and
\be\label{ab}
a\in(p_{k+1},p_k],\qquad
b\in[p_{m+1},p_m).
\ee

The $\eps$-clipped vector (the flattest $\eps$-approximation~\cite{HY2010,HD2021,HOS2018}) is $\upb^{(\eps)}\in\prob^\da(d)$ defined by
\be\label{clipped}
\underline{p}_x^{(\eps)}=
\begin{cases}
a & x\in[k]\\
p_x & k<x\le m\\
b & x\in\{m+1,\ldots,d\}.
\end{cases}
\ee
Equivalently,
\be\label{justclip}
\underline{p}_x^{(\eps)}=\max\{b,\min\{a,p_x\}\}.
\ee

In this work we will also use extremal elements with respect to \emph{relative majorization}.  
For $\p,\q\in\prob(d)$ and $\p',\q'\in\prob(d')$ we say that $(\p,\q)$ relatively majorizes $(\p',\q')$, written
\be
(\p,\q)\succ(\p',\q'),
\ee
if there exists a $d'\times d$ column-stochastic matrix $E$ such that
\be
E\p=\p',\qquad E\q=\q'.
\ee
We write $(\p,\q)\sim(\p',\q')$ if both directions hold.

Relative majorization has appeared under several names, including $d$-majorization~\cite{Veinott1971}, thermo-majorization~\cite{OH2013}, and matrix majorization~\cite{Dahl1999}.  
It admits an elegant geometric characterization in terms of testing regions (see~\cite[Ch.~4]{Gour2025}). A useful connection to majorization occurs when $\q$ has rational components,
\be\label{rational}
\q=\left(\frac{k_1}{k},\ldots,\frac{k_d}{k}\right)^T,\qquad
k_x\in\mbb{N}\;,\qquad k=\sum_{x\in[d]} k_x .
\ee
For $\p\in\prob(d)$ define
\ba\label{directsum}
\t &\eqdef \bigoplus_{x\in[d]} p_x \u^{(k_x)}\\
&=\Big(\underbrace{\tfrac{p_1}{k_1},\ldots,\tfrac{p_1}{k_1}}_{k_1},
\underbrace{\tfrac{p_2}{k_2},\ldots,\tfrac{p_2}{k_2}}_{k_2},
\ldots,
\underbrace{\tfrac{p_d}{k_d},\ldots,\tfrac{p_d}{k_d}}_{k_d}\Big)^T .
\ea
Then~\cite[Sec.~4.3]{Gour2025}
\be
(\p,\q)\sim(\t,\u),
\ee
where $\u$ is the uniform distribution in $\prob(k)$.  
Consequently any classical divergence $\D$ satisfies $\D(\p\|\q)=\D(\t\|\u)$.

\subsection{Classical Divergences}

A function acting on pairs of probability vectors in all finite dimensions
\be\label{dbd}
\D:\bigcup_{d\in\mbb{N}}\Big\{\prob(d)\times\prob(d)\Big\}\to\mbb{R}\cup\{\infty\}
\ee 
is called a \emph{classical divergence} if it satisfies the data processing inequality (DPI): for every $d,d'\in\mbb{N}$, $\p,\q\in\prob(d)$, and $d'\times d$ column stochastic matrix $E$, we have 
\be
\D\big(E\p\big\|E\q\big)\leq \D(\p\|\q)\;.
\ee
In other words, a classical divergence is a function on pairs of probability vectors that is monotone under relative majorization.

We follow the terminology of~\cite{GT2021} and call a divergence $\D$ a \emph{relative entropy} if, in addition, it is additive under tensor products,
\be
\D(\p\otimes\p'\|\q\otimes\q')=\D(\p\|\q)+\D(\p'\|\q')\;,
\ee
and is normalized such that $\D(\e_1\|\u)=1$, where $\e_1=(1,0)$ and $\u=(\tfrac12,\tfrac12)$. 
A fundamental family of examples is given by the R\'enyi divergences: for $\alpha\in(0,1)\cup(1,\infty)$ and $\p,\q\in\prob(d)$,
\be
D_\alpha(\p\|\q)\eqdef \frac{1}{\alpha-1}\log\sum_{x\in[d]}p_x^\alpha q_x^{1-\alpha}\;,
\ee
with the convention that $D_\alpha(\p\|\q)=\infty$ if $\supp(\p)\not\subseteq\supp(\q)$ for $\alpha>1$, and with the continuous extensions at $\alpha=1$, $\alpha=0$, and $\alpha=\infty$, given by the Kullback--Leibler divergence, the min-relative entropy, and the max-relative entropy respectively.
We use base-two logarithms, with $\log0=-\infty$. Terms with $p_x=q_x=0$ are omitted; a likelihood ratio with $p_x>0=q_x$ is $+\infty$, while $0/0$ may be assigned zero for sorting. At order zero,
\be
D_0(\p\|\q)=-\log\sum_{x:p_x>0}q_x,\qquad
H_0(\p)=\log|\supp(\p)|.
\ee
Thus $p_x^0$ means the indicator of $p_x>0$ in order-zero power sums. At order one we use $0\log0=0$, and $D_1(\p\|\q)=\infty$ if $\supp(\p)\not\subseteq\supp(\q)$. For $0<\alpha<1$, a zero power sum gives $D_\alpha=\infty$. These extended-real conventions also apply to the support cases of the corresponding quantum formulas.

Recently it was shown~\cite{MPS+2021} that any classical divergence $\D$ that is continuous in its second argument is a relative entropy if and only if it can be written as a convex combination (i.e., an integral mixture) of R\'enyi divergences.

Given $\eps\in[0,1]$ and a classical divergence $\D$ (not necessarily a relative entropy), we define the $\eps$-smoothed variant of $\D$ by
\be\label{csmoothed}
\D^\eps(\p\|\q)\eqdef \min_{\p'\in\mb^\eps(\p)} \D(\p'\|\q)\;.
\ee
The smoothing reflects an operational tolerance allowing $\p$ to be approximated within accuracy~$\eps$ before evaluating the divergence.
Importantly, the smoothed divergence $\D^\eps$ is itself a classical divergence, i.e., it continues to satisfy the data processing inequality.
This follows from the monotonicity of both the feasible set $\mb^\eps(\p)$ under stochastic maps and of $\D$ itself.
By contrast, if smoothing were defined via a maximization over $\mb^\eps(\p)$ rather than a minimization, the DPI would generally fail, since maximization is incompatible with the monotonicity required under data processing.

So far we only considered standard smoothing of R\'enyi divergences. Another key smoothed divergence widely used in quantum information is the hypothesis-testing divergence. For any $\p,\q\in\prob(d)$ and $\eps\in[0,1)$,
the \emph{classical} hypothesis-testing divergence is defined as
\be\label{tstv}
D_{H}^\eps(\p\|\q)\eqdef-\log\min\big\{\q\cdot\t\;:\;\p\cdot\t\geq 1-\eps\;,\;\t\in[0,1]^d\big\}.
\ee
The hypothesis-testing divergence admits the following closed form. Without loss of generality assume
\be\nonumber
\frac{p_1}{q_1}\ge\cdots\ge\frac{p_d}{q_d}.
\ee
Set
\be
a_k\eqdef\sum_{x\in[k]}p_x\;,\qquad 
b_k\eqdef\sum_{x\in[k]}q_x\;,\qquad 
\forall\;k\in\{0,1,\ldots,d\}.
\ee
Then (see, e.g., Sec.~8.6 of~\cite{Gour2025})
\be\label{formula}
D^\eps_{H}\left(\p\big\|\q\right)
=-\log\left(b_\ell+\frac{q_{\ell+1}}{p_{\ell+1}}(1-\eps-a_{\ell})\right)
\ee
where $\ell\in\{0,\ldots,d-1\}$ satisfies $a_\ell<1-\eps\le a_{\ell+1}$.  
For $\q=\u$ this reduces to
\be\label{dhd}
D^\eps_{H}\left(\p\big\|\u\right)
=\log(d)-\log\left(\ell+\frac{1}{p_{\ell+1}}(1-\eps-\|\p\|_{(\ell)})\right)
\ee 
where $\ell\in\{0,\ldots,d-1\}$ satisfies $\|\p\|_{(\ell)}<1-\eps\le \|\p\|_{(\ell+1)}$.

\subsection{Quantum Divergences}

We use $A$, $B$, and $R$ to denote both quantum systems (or registers) and their associated Hilbert spaces. The set of density operators on $A$ is denoted by $\md(A)$, and the set of completely positive trace-preserving (CPTP) maps from $A$ to $B$ is denoted by $\cptp(A \to B)$. The maximally mixed (uniform) state in $\md(A)$ is written as $\u_A$.
For $\rho,\sigma \in \md(A)$ the primary metric we use is the trace distance,
\be
T(\rho,\sigma)\eqdef\frac12\|\rho-\sigma\|_1\;.
\ee
The $\eps$-ball around $\rho\in\md(A)$ is
\be\label{epsball}
\mb^\eps(\rho) \eqdef \left\{\sigma\in\md(A) \;:\; T(\rho,\sigma) \leq \eps \right\}.
\ee

Consider a function acting on pairs of quantum states in all finite dimensions:
\be\label{quantum-divergence}
\D:\bigcup_{A}\Big\{\md(A)\times\md(A)\Big\}\to\mbb{R}\cup\{\infty\}.
\ee
The function $\D$ is called a \emph{quantum divergence} if it satisfies the Data Processing Inequality (DPI): for every $\mE\in\cptp(A\to B)$ and $\rho,\sigma\in\md(A)$,
\be
\D\big(\mE(\rho)\big\|\mE(\sigma)\big)\leq \D(\rho\|\sigma)\;.
\ee
As in the classical case, a quantum divergence $\D$ is called a \emph{relative entropy} if in addition it is additive under tensor products and is normalized such that $\D(|0\lr 0|\|\frac12I_2)=1$. 

The minimal quantum extension of a classical divergence $\D$ is the smallest quantum divergence that reduces to $\D$ on classical (i.e., commuting) states. It is well known (e.g.~\cite{Tomamichel2015,KW2024,Gour2025}) that this extension coincides with the measured quantum divergence defined for every $\rho,\sigma\in\md(A)$ by
\be\label{measured}
\D^{\mathbb{M}}(\rho\|\sigma)\eqdef\sup_{\mE\in\cptp(A\to X)}\D\big(\mE(\rho)\big\|\mE(\sigma)\big)
\ee
where the supremum is taken over all POVM channels $\mE\in\cptp(A\to X)$ and over all classical systems $X$.

If $\D$ is a classical relative entropy, its minimal extension $\D^{\mathbb{M}}$ is not necessarily additive. However, in the important case $\D=D_\alpha$, the R\'enyi divergence of order $\alpha\in[0,\infty]$, one can obtain a quantum relative entropy by regularizing $D_\alpha^{\mathbb{M}}$. Specifically (see, e.g.~\cite{Tomamichel2015,Gour2025} and references therein),
\be
\lim_{n\to\infty}\frac1nD_\alpha^{\mathbb{M}}\left(\rho^{\otimes n}\big\|\sigma^{\otimes n}\right)=\widetilde D_{\alpha}(\rho\|\sigma)\;,
\ee
Here $\widetilde D_\alpha$ denotes the regularized measured R\'enyi divergence. For $\alpha\ge1/2$ it coincides with the usual sandwiched R\'enyi divergence~\cite{MDS+2013,WWY2014}. Below $1/2$ it is the swapped-order extension described below, rather than the usual sandwiched formula at the same order~\cite{MH2022}. For $\alpha\ne1,\infty$,
\be
\widetilde D_{\alpha}(\rho\|\sigma)=
	\begin{cases}
	\frac1{\alpha-1}\log \widetilde Q_\alpha(\rho\|\sigma)
	& \text{if }\substack{\frac12\leq\alpha<1 \text{ and }\rho\not\perp\sigma\\\text{ or }\alpha>1 \text{ and }\rho\ll\sigma} \\
	\frac1{\alpha-1}\log \widetilde Q_{1-\alpha}(\sigma\|\rho)
	& \text{if }0\leq \alpha<\frac12 \text{ and }\rho\not\perp\sigma\\
	\infty & \text{otherwise.}
	\end{cases}
\ee
Here $\rho \ll \sigma$ indicates that the support of $\rho$ is contained in that of $\sigma$, while $\rho \not\perp \sigma$ means $\tr[\rho \sigma] \neq 0$. The quantity $\widetilde Q_\alpha(\rho\|\sigma)$ is defined as
\be
\tQ_\alpha(\rho\|\sigma)\eqdef\tr\left(\sigma^{\frac{1-\alpha}{2\alpha}}\rho\sigma^{\frac{1-\alpha}{2\alpha}}\right)^\alpha\;.
\ee
Thus $\tD_\alpha$ is the smallest additive divergence that reduces on classical states to the R\'enyi divergence of order $\alpha$.
More explicitly, if $D_\gamma^{\rm sw}$ denotes the usual sandwiched divergence, then
\be
\widetilde D_\alpha(\rho\|\sigma)
=\frac{\alpha}{1-\alpha}D_{1-\alpha}^{\rm sw}(\sigma\|\rho),
\qquad 0<\alpha<\frac12.
\ee
At $\alpha=0$ the definition gives $\widetilde D_0(\rho\|\sigma)=-\log\tr(\Pi_\rho\sigma)$.

For $\alpha=1$, the sandwiched R\'enyi relative entropy reduces to the Umegaki relative entropy:
\be 
D(\rho\|\sigma)\eqdef\tr[\rho\log(\rho)]-\tr[\rho\log(\sigma)]\;.
\ee
For $\alpha=\infty$, it reduces to the max-relative entropy:
\be
D_{\max}(\rho\|\sigma)\eqdef\inf_{t\in\mbb{R}_+}\big\{\log(t)\;:\;t\sigma\geq\rho\big\}\;.
\ee

Smoothed entropic quantities play a central role in single-shot quantum information theory, as they characterize optimal rates of quantum information-processing tasks. Given $\eps\in(0,1)$ and a quantum divergence $\D$, we define the $\eps$-smoothed divergence by
\be\label{smoothed}
\D^\eps(\rho\|\sigma)\eqdef \inf_{\rho'\in\mb^\eps(\rho)} \D(\rho'\|\sigma)\;.
\ee
If $\D(\cdot\|\sigma)$ is lower semicontinuous, compactness of $\mb^\eps(\rho)$ ensures that the infimum is a minimum.
In particular, the smoothed max-relative entropy
\be
D_{\max}^\eps(\rho\|\sigma)\eqdef\min_{\rho'\in\mb^\eps(\rho)}D_{\max}\left(\rho'\|\sigma\right)
\ee
is a fundamental quantity in quantum information theory and appears in numerous single-shot applications.
We emphasize that smoothing is performed with respect to the trace distance. While other choices, such as the purified distance, are common in the literature, the trace distance is crucial for our analysis. 

Another important smoothed divergence arises from the information spectrum divergence. Consider the following two variants introduced in~\cite{DL2015}. For $\rho,\sigma\in\md(A)$ and $\eps\in(0,1)$,
\ba\label{iss}
\underline{D}_s^\eps(\rho\|\sigma)&\eqdef\sup_{\lambda\in\mbb{R}}\{\lambda\;:\;\tr(\rho-2^\lambda\sigma)_+\geq1-\eps\}\\
\overline{D}_s^\eps(\rho\|\sigma)&\eqdef\inf_{\lambda\in\mbb{R}}\{\lambda\;:\;\tr(\rho-2^\lambda\sigma)_+\leq\eps\}
\ea
where $(X)_+$ and $(X)_-$ denote the positive and negative parts of a Hermitian matrix $X$. These divergences are variants of the definition originally introduced in~\cite{TH2013}.

Let $\delta=\tr(\rho\Pi_{\ker\sigma})$. The continuous function $\tr(\rho-2^\lambda\sigma)_+$ decreases from $1$ to $\delta$ and is strictly decreasing wherever its value exceeds $\delta$. Consequently, if $\delta<1-\eps$, in particular if $\rho\ll\sigma$, then
\be
\underline{D}_s^\eps(\rho\|\sigma)=\overline{D}_s^{1-\eps}(\rho\|\sigma)
\ee
(see~\cite{DL2015}). For singular reference states this identity must be qualified: a plateau at the limiting value can give different endpoints when $\delta=1-\eps$. We use $\overline D_s^\eps$ directly for arbitrary supports, with $\inf\emptyset=\infty$. Following the notation of~\cite{RLD2025} we define
\be
\widetilde{D}_{\max}^\eps(\rho\|\sigma)\eqdef\overline{D}_s^\eps(\rho\|\sigma)\;.
\ee
This notation reflects the fact that $\overline{D}_s^\eps$ is a smoothed variant of $D_{\max}$; in particular $\overline{D}_s^0=D_{\max}$.

Moreover,~\cite{RLD2025} showed that $\widetilde{D}_{\max}^\eps$ is related to the hypothesis-testing divergence through
\ba\label{is}
\widetilde{D}_{\max}^\eps(\rho\|\sigma)
=\sup_{\delta\in(\eps,1]}\left\{D_{H}^{1-\delta}(\rho\|\sigma)+\log(\delta-\eps)\right\}\\
D_{H}^{1-\eps}(\rho\|\sigma)
=\inf_{\delta\in[0,\eps)}\left\{\widetilde{D}_{\max}^{\delta}(\rho\|\sigma)-\log(\eps-\delta)\right\}\;,
\ea
where the hypothesis-testing divergence is defined as
\be
D_{H}^{\eps}(\rho\|\sigma)\eqdef-\log
\min\Big\{\tr[\Lambda\sigma]\;:\;\tr[\Lambda\rho]\geq 1-\eps\;,\;0\leq\Lambda\leq I_A\Big\}.
\ee

\section{Smoothing, Relative Majorization, and Universal Bounds}\label{Sec4}

In this section we develop the structural framework underlying the universal bounds. We first construct the flattest approximation of a probability vector relative to a fixed reference distribution. This clipped vector minimizes every classical divergence over the total-variation ball, including when the reference distribution has zero components.

We then distinguish smoothing a measured divergence from taking the measured extension of a classically smoothed divergence. This distinction identifies the quantum bounds that follow from a classical optimization. Finally, we prove the classical reduction and, under a regularity assumption, the further reduction to a uniform reference distribution. The theorem proofs use these results directly; in particular, the split--merge construction in Theorems~\ref{thmg1} and~\ref{thmg1sub} reduces their optimizations to binary pairs.

\subsection{Flattest $\eps$-Approximations under Relative Majorization}\label{sec:arm}

We begin by showing that a minimal element of $\mb^\eps(\p)$ exists also with respect to relative majorization.

\bmyd
Let $\p,\q\in\prob(d)$ and $\eps\in[0,1]$. The flattest $\eps$-approximation of $\p$ relative to $\q$ is a vector $\p^{(\eps)}\in\mb^\eps(\p)$ such that
\be
(\p',\q)\succ(\p^{(\eps)},\q)\qquad\forall\;\p'\in\mb^\eps(\p)\;.
\ee
\emyd

In the next theorem we show that the flattest $\eps$-approximation of $\p$ relative to $\q$ always exists and is given by a clipped vector. The two clipping thresholds admit a direct operational characterization in terms of two variants of the smoothed max-relative entropy. Specifically, we define
\begin{align}
D_{\max}^{\eps,+}(\p\|\q)
&\coloneqq
\log_2 \inf\left\{\lambda>0:\;\|\p-\lambda \q\|_+ \le \eps \right\},
\\
D_{\max}^{\eps,-}(\p\|\q)
&\coloneqq
\log_2 \sup\left\{\lambda\ge0:\;\|\p-\lambda \q\|_- \le \eps \right\},
\end{align}
where $\|\cdot\|_+$ and $\|\cdot\|_-$ denote the total mass of the positive and negative parts, respectively. The upper cutoff of the likelihood ratio is given by
\be
Q_{\max}^{\eps,+}(\p\|\q)\eqdef2^{D_{\max}^{\eps,+}(\p\|\q)}\;,
\ee
while the lower cutoff is given by
\be
Q_{\max}^{\eps,-}(\p\|\q)\eqdef2^{D_{\max}^{\eps,-}(\p\|\q)}\;.
\ee
These quantities quantify, respectively, how much the distribution $\p$ can locally exceed or fall below a scaled version of $\q$ under the allowed total-variation perturbation.

For strictly positive $\q$ and $0<\eps<\frac12\|\p-\q\|_1$, the clipped likelihood ratio lies in the bounded interval $[b,a]$ defined below. If $\q$ has zero components, the upper cutoff may be infinite; the construction below includes this case. This two-sided clipping reflects the fact that total-variation smoothing permits a limited redistribution of probability mass in both directions and therefore enforces simultaneous control over overly distinguishable and overly negligible events. Conceptually, this structure is reminiscent of thresholding phenomena in robust hypothesis testing, where least-favorable distributions arise from truncating likelihood ratios at both extremes (see, e.g.,~\cite{Huber1965,HuberStrassen1973,Poor1994}).

The smoothed divergence $D_{\max}^{\eps,+}$ has a quantum analogue originally defined in~\cite{DL2015,TH2013} and studied further in~\cite{RLD2025} under the notation $\widetilde{D}_{\max}^\eps$. For classical probability vectors, its relation to max-relative entropy smoothed over normalized distributions is
\be
D_{\max}^{\eps}(\p\|\q)
=\max\{0,D_{\max}^{\eps,+}(\p\|\q)\}
=\inf_{\p'\in\mb^\eps(\p)}D_{\max}(\p'\|\q).
\ee
In particular, the two quantities coincide when $\eps<\frac12\|\p-\q\|_1$.
For $q_x>0$ for every $x\in[d]$, both $Q_{\max}^{\eps,\pm}$ admit the following closed forms. Set $r_x\eqdef p_x/q_x$ for all $x\in[d]$ and suppose w.l.o.g.\ that $r_1\geq r_2\geq\cdots\geq r_d$. Then,
\be\label{aaa}
Q_{\max}^{\eps,+}(\p\|\q)=\max_{m\in[d]}\frac{\sum_{x\in[m]}p_x-\eps}{\sum_{x\in[m]}q_x}\;,
\ee
and
\be
Q_{\max}^{\eps,-}(\p\|\q)=\min_{\ell\in[d]}\frac{\sum_{x=\ell}^dp_x+\eps}{\sum_{x=\ell}^dq_x}\;.
\ee

Let $\eps\in[0,1]$ and $\p,\q\in\prob(d)$.
If $\eps=0$, set $\p^{(\eps)}=\p$.
If $\frac12\|\p-\q\|_1\le\eps$, set $\p^{(\eps)}=\q$.
Otherwise, suppose
\be
0<\eps<\frac12\|\p-\q\|_1,
\ee
and define
\be
\delta\eqdef\sum_{x:q_x=0}p_x.
\ee
Let
\be
a\eqdef Q_{\max}^{\eps,+}(\p\|\q),
\qquad
b\eqdef Q_{\max}^{\eps,-}(\p\|\q),
\ee
using the positive- and negative-part characterizations
\be
a=\inf\Big\{\lambda>0:
\sum_{x\in[d]}(p_x-\lambda q_x)_+\le\eps\Big\},
\ee
and
\be
b=\sup\Big\{\lambda\ge0:
\sum_{x\in[d]}(\lambda q_x-p_x)_+\le\eps\Big\}.
\ee
Here $\inf\emptyset=\infty$.
In particular, $a>1>b>0$, and $a=\infty$ if and only if
$\eps<\delta$.

For every $x$ with $q_x>0$, set $r_x\eqdef p_x/q_x$.
Define the components of $\p^{(\eps)}$ by
\be\label{pstar}
p_x^{(\eps)}=
\begin{cases}
q_x\max\{b,\min\{a,r_x\}\}
& \text{if }q_x>0,\\[2pt]
\left(1-\frac{\eps}{\delta}\right)p_x
& \text{if }q_x=0\text{ and }\eps<\delta,\\[2pt]
0
& \text{if }q_x=0\text{ and }\eps\ge\delta.
\end{cases}
\ee
When $a=\infty$, the first case is understood as
$p_x^{(\eps)}=\max\{p_x,bq_x\}$.

\medskip
\noindent\textbf{Theorem~\ref{thmg0} (structural formulation).}
{\it
The vector $\p^{(\eps)}$ defined above is the flattest
$\eps$-approximation of $\p$ relative to $\q$.
Consequently, every classical divergence $\D$ satisfies
\be
\D^\eps(\p\|\q)
=
\D\big(\p^{(\eps)}\big\|\q\big).
\ee
}

This structural description of smoothing in terms of clipped vectors clarifies the classical geometry of the $\eps$-ball. We now turn to the quantum setting and examine how smoothing interacts with the passage between classical and quantum domains. In particular, we must distinguish between two natural procedures: smoothing before lifting to the quantum domain, and smoothing after taking the minimal quantum extension.

\subsection{Two Types of Smoothing in the Quantum Domain}

There are two natural ways to define a smoothed quantum divergence associated with a classical divergence $\D$. One may first lift $\D$ to its minimal quantum extension and then smooth in the quantum domain, or alternatively smooth in the classical domain and then lift to the quantum domain. While these two procedures coincide for commuting states, they differ in general. In this subsection we compare these constructions and clarify their relationship.

Let $\D$ be a classical divergence (not necessarily a relative entropy), and let $\D^{\mathbb{M}}$ be its minimal quantum extension as defined in~\eqref{measured}. Following~\eqref{smoothed}, its smoothed version is defined as
\be\label{meps}
\D^{\mathbb{M},\eps}(\rho\|\sigma)\eqdef\inf_{\trho\in\mb^\eps(\rho)}\D^{\mathbb{M}}(\trho\|\sigma)\;.
\ee
One can also define an alternative smoothed quantum divergence by first applying smoothing in the classical domain and only then extending it to the quantum domain via measurements. The resulting quantity is
\be\label{epsm}
\D^{\eps,\mathbb{M}}(\rho\|\sigma)\eqdef\sup_{\mE\in\cptp(A\to X)}\D^\eps\big(\mE(\rho)\big\|\mE(\sigma)\big)\;.
\ee
The difference between these two divergences lies in the order in which smoothing and the lift to the quantum domain are applied. Accordingly, the superscript $\mathbb{M},\eps$ denotes first lifting to the quantum domain and then smoothing, while $\eps,\mathbb{M}$ denotes first smoothing in the classical domain and then lifting via measurement.

The following lemma shows that these two divergences coincide for commuting states. Although the result is essentially immediate, a subtlety must be addressed. On the right-hand side of~\eqref{epsm}, the smoothed divergence $\D^\eps$ is classical, and therefore the smoothing is performed only over diagonal states within the classical $\eps$-ball around $\rho$. In contrast, in~\eqref{meps}, even if both $\rho$ and $\sigma$ are diagonal, the smoothing is carried out over all (possibly non-diagonal) states $\trho\in\mb^\eps(\rho)$. One must therefore show that restricting the smoothing to diagonal states leaves the infimum unchanged.

\bmyl
Let $\rho,\sigma\in\md(A)$. Then,
\be
\D^{\mathbb{M},\eps}(\rho\|\sigma)\geq \D^{\eps,\mathbb{M}}(\rho\|\sigma)\;,
\ee
with equality if $\rho$ and $\sigma$ commute.
\emyl

\begin{proof}
By definition,
\ba\label{62}
\D^{\mathbb{M},\eps}(\rho\|\sigma)
&=\inf_{\trho\in\mb^\eps(\rho)}\sup_{\mE\in\cptp(A\to X)}\D\big(\mE(\trho)\big\|\mE(\sigma)\big)\\
&\geq \sup_{\mE\in\cptp(A\to X)}\inf_{\trho\in\mb^\eps(\rho)}\D\big(\mE(\trho)\big\|\mE(\sigma)\big)\\
&\geq \sup_{\mE\in\cptp(A\to X)}\D^\eps\big(\mE(\rho)\big\|\mE(\sigma)\big)\\
&=\D^{\eps,\mathbb{M}}(\rho\|\sigma)\;,
\ea
where the first inequality follows from the weak minimax inequality, and the second from the fact that $\mE(\trho)$ is $\eps$-close to $\mE(\rho)$ by the data processing inequality, since $\trho$ is $\eps$-close to $\rho$.

To prove equality, assume that $\rho$ and $\sigma$ commute and denote by $\mb_c^\eps(\rho)$ the set of all classical (i.e.\ diagonal) states that are $\eps$-close to $\rho$. Since $\mb_c^\eps(\rho)\subseteq \mb^\eps(\rho)$, restricting the minimization in~\eqref{62} to $\mb_c^\eps(\rho)$ yields
\ba
\D^{\mathbb{M},\eps}(\rho\|\sigma)
&\leq\inf_{\trho\in\mb^\eps_c(\rho)}\sup_{\mE\in\cptp(A\to X)}\D\big(\mE(\trho)\big\|\mE(\sigma)\big)\\
&\leq\inf_{\trho\in\mb^\eps_c(\rho)}\D\big(\trho\big\|\sigma\big)\\
&=\D^\eps(\rho\|\sigma)\;.
\ea
Since the opposite inequality was shown in~\eqref{62}, we conclude that
$\D^{\mathbb{M},\eps}(\rho\|\sigma)=\D^{\eps,\mathbb{M}}(\rho\|\sigma)$ whenever $\rho$ and $\sigma$ commute.
\end{proof}

Having clarified the structural role of smoothing and its behavior under measurement, we now turn to the problem of optimal universal bounds between divergences. Our goal is to quantify, in a dimension-independent manner, how far a smoothed divergence can deviate from a reference divergence. We formalize this problem below and show that the resulting optimization can be reduced to a classical one.

\subsection{Universal Bounds and Their Reduction to the Classical Domain}

We begin by introducing the notion of optimal universal bounds between two divergences. Given two quantum divergences, we define dimension-independent constants that quantify the maximal possible gap between them, both in the upper and lower directions. Although these definitions involve an optimization over arbitrary quantum systems, we show that in many cases the problem reduces entirely to classical probability distributions, and can even be further reduced to distributions relative to the uniform state.

We use $\D$ and $\Ediv$ for generic divergences, reserving symbols such as $\widetilde Q_\alpha$ and $Q_{\max}^{\eps,\pm}$ for the corresponding power quantities. Let $\D$ and $\Ediv$ be two quantum divergences, $\varepsilon\in(0,1)$, and $\D^\eps$ a smoothed variant of $\D$. We seek dimension-independent constants that quantify the maximal possible deviation between the smoothed divergence $\D^\varepsilon$ and the reference divergence $\Ediv$. Specifically,  we consider bounds of the form (for all $\rho,\sigma\in\md(A)$)
\be
\Ediv(\rho\|\sigma)-\nu(\eps)\leq\D^\eps(\rho\|\sigma)\leq \Ediv(\rho\|\sigma)+\mu(\eps)\;,
\ee
where $\mu(\eps),\nu(\eps)\in\mbb{R}$ are constants independent of the dimension of $A$. We say that such relations are \emph{optimal} if
\begin{align}
\mu(\eps)&=\sup_A\sup_{\rho,\sigma\in\md(A)}\Big\{\D^\eps(\rho\|\sigma)- \Ediv(\rho\|\sigma)\Big\}\label{mu}\\
\nu(\eps)&=\sup_A\sup_{\rho,\sigma\in\md(A)}\Big\{\Ediv(\rho\|\sigma)-\D^\eps(\rho\|\sigma)\Big\}\label{nu}\;,
\end{align}
where the supremum is over all finite-dimensional systems $A$.

By definition, $\mu(\varepsilon)=\infty$ indicates that no finite dimension-independent upper bound exists, and similarly $\nu(\varepsilon)=\infty$ means that no finite lower bound holds uniformly over all systems.
If both $\Ediv$ and $\D^\varepsilon$ are normalized divergences, i.e., they vanish whenever $\rho=\sigma$, then evaluating the bounds at $\rho=\sigma$ immediately yields
\be
\mu(\varepsilon)\ge 0
\qquad\text{and}\qquad
\nu(\varepsilon)\ge 0\;.
\ee
In contrast, for divergences that are not normalized, such as the hypothesis-testing divergence or $\widetilde D_{\max}^{\varepsilon}$, the optimal constants $\mu(\varepsilon)$ and $\nu(\varepsilon)$ may be negative, as will be seen below.

In general, computing $\mu(\eps)$ and $\nu(\eps)$ appears challenging, since it involves an optimization over pairs of density matrices of arbitrarily large dimension. However, we now show that in many cases the expressions for $\mu(\eps)$ and $\nu(\eps)$ given in~\eqref{mu} and~\eqref{nu} can be simplified significantly. In particular, many optimal universal bounds follow directly from their classical counterparts.

Let $\eps\in(0,1)$, $\D$ be a \emph{classical} divergence, and $\Ediv$ a \emph{quantum} divergence. Define
\ba
\kappa\eqdef\sup_A\sup_{\rho,\sigma\in\md(A)}\left\{\D^{\mathbb{M}}(\rho\|\sigma)- \Ediv(\rho\|\sigma)\right\}\;,
\ea
where $\D^\mathbb{M}$ is the minimal (i.e.\ measured) quantum extension of $\D$. Then:

\bmyl\label{lemc}
 \be\label{kappa}
 \kappa=\sup_{d\in\mbb{N}}\sup_{\p,\q\in\prob(d)}\left\{\D(\p\|\q)- \Ediv(\p\|\q)\right\}\;.
 \ee
\emyl

\begin{proof}
By definition, the right-hand side of~\eqref{kappa} is no greater than $\kappa$. To prove the opposite inequality, let $\rho,\sigma\in\md(A)$. Then,
\ba
&\D^{\mathbb{M}}(\rho\|\sigma)- \Ediv(\rho\|\sigma)\\
&=\sup_{\mE\in\cptp(A\to X)}\D\big(\mE(\rho)\big\|\mE(\sigma)\big) -\Ediv(\rho\|\sigma)\\
&\leq \sup_{\mE\in\cptp(A\to X)}\Big\{\D\big(\mE(\rho)\big\|\mE(\sigma)\big) -\Ediv\big(\mE(\rho)\big\|\mE(\sigma)\big)\Big\}\\
&\leq \sup_{d\in\mbb{N}}\sup_{\p,\q\in\prob(d)}\Big\{\D(\p\|\q)- \Ediv(\p\|\q)\Big\}\;.
\ea
Since this inequality holds for all $A$ and all $\rho,\sigma\in\md(A)$, we conclude that $\kappa$ is no greater than the right-hand side of~\eqref{kappa}. This completes the proof.
\end{proof}

The direction of this reduction is essential: the positively occurring term must be the measured extension $\D^{\mathbb{M}}$, while the subtracted term $\Ediv$ may be any quantum divergence satisfying DPI. Measurement can discard distinguishability; DPI controls this loss in the subtracted term, but does not restore it in a positively occurring nonmeasured divergence. Thus the lemma yields lower bounds on $D_H^\eps$ in terms of $D_\alpha^{\mathbb{M}}$, but does not by itself replace that comparator by the generally larger Petz divergence $D_\alpha$ for $0<\alpha<1$. The latter replacement would strengthen the bound and needs an independent quantum argument. In Section~II.C we therefore invoke the known Petz bound~\eqref{oub}; the classical constructions in the proof of Theorem~\ref{thmg4} then establish sharpness of its correction term for $\eps\le\alpha<1$. Theorem~\ref{thmg3}, in contrast, concerns the upper bound on $D_H^\eps$.

Using the vector in~\eqref{directsum} we can simplify the expression for $\kappa$ even further.

\bmyc\label{cor001}
Assume that, for every $d$ and fixed $\p\in\prob(d)$, $\Ediv(\p\|\q)$ is finite whenever $\q$ has full support, and that
$\q\mapsto\D(\p\|\q)-\Ediv(\p\|\q)$ is lower semicontinuous at every $\q$ for which $\Ediv(\p\|\q)<\infty$ (on this finite-$\Ediv$ domain). Then
\be\label{cor800}
\kappa=\sup_{d\in\mbb{N}}\sup_{\p\in\prob(d)}\Big\{\D(\p\|\u)-\Ediv(\p\|\u)\Big\}\;.
\ee
\emyc

\begin{proof}
Let $K$ denote the right-hand side of~\eqref{cor800}. Fix $\p,\q\in\prob(d)$ with $\Ediv(\p\|\q)<\infty$, and choose strictly positive rational probability vectors $\q^{(j)}\to\q$. Lower semicontinuity gives
\be
\D(\p\|\q)-\Ediv(\p\|\q)
\le\liminf_{j\to\infty}\big\{\D(\p\|\q^{(j)})-\Ediv(\p\|\q^{(j)})\big\}.
\ee
For every $j$, the embedding~\eqref{directsum} gives
$(\p,\q^{(j)})\sim(\t^{(j)},\u)$, possibly in a larger dimension.
DPI in both directions preserves each divergence, so each difference on the right is at most $K$. Lemma~\ref{lemc} now implies $\kappa\le K$. The converse follows by restricting the reference distribution to $\u$.
\end{proof}

The regularity assumption holds for the classical quantities used below, including their boundary cases. For fixed $\p$, the R\'enyi divergences, their normalized smoothed versions, the subnormalized smoothed versions used here, and $D_H^\eps$ are continuous in $\q$ at finite values and converge to $+\infty$ at infinite values. For positive R\'enyi order, lower semicontinuity of a smoothed quantity follows from joint lower semicontinuity and compactness of its $\q$-independent smoothing set. For the opposite bound, if $\q^{(j)}\to\q$ has full support, then $\q^{(j)}\ge(1-\eta_j)\q$ componentwise for some $\eta_j\to0$, and
\be
D_\alpha(\r\|\q^{(j)})
\le D_\alpha(\r\|\q)-\log(1-\eta_j).
\ee
Evaluating at a minimizer for $\q$ gives the same upper bound after smoothing. At order zero, use
$D_0(\p\|\q)=-\log\sum_{x:p_x>0}q_x$ and $D_0^\eps(\p\|\q)=0$ for $\eps>0$.
For $D_H^\eps$, the minimum in~\eqref{tstv} is a continuous function of $\q$, since it minimizes a linear function over a fixed compact set. Consequently, subtracting a quantity with a finite limit preserves the required lower semicontinuity. This verifies the assumption for each application of Corollary~\ref{cor001} below; DPI alone would not suffice.

Finally, if $\D$ and $\Ediv$ are relative entropies (i.e.\ additive divergences), then for every $\p\in\prob(d)$ we can define the entropy functions
\be
H_{{}_{\!\D}}(\p)\eqdef\log(d)-\D(\p\|\u)
\ee
and
\be
 H_{{}_{\!\Ediv}}(\p)\eqdef\log(d)-\Ediv(\p\|\u)\;.
\ee
In terms of these functions we can express $\kappa$ as
\be\label{newsup}
\kappa=\sup_{d\in\mbb{N}}\sup_{\p\in\prob^\da(d)}\Big\{ H_{{}_{\!\Ediv}}(\p)-H_{{}_{\!\D}}(\p)\Big\}\;.
\ee

The classical reduction above applies to an arbitrary divergence $\D$, including a smoothed divergence; the further uniform-reference reduction requires the regularity assumption in Corollary~\ref{cor001}. We now apply these reductions to the main theorems, using the clipped optimizer and direct arguments tailored to each bound.

\section{Proof of Theorem~\ref{thmg1}}\label{Sec5}

\noindent\textbf{Theorem 1.} {\it Let  $\eps\in(0,1)$, $\alpha,\beta\in[0,\infty]$, 
\be\label{mainex-restated}
\mu(\eps,\alpha,\beta)\eqdef\sup_A\sup_{\rho,\sigma\in\md(A)}\Big\{D_\beta^{\eps,\mathbb{M}}(\rho\|\sigma)- \D_\alpha(\rho\|\sigma)\Big\}
\ee
and
\be
\theta\eqdef
\begin{cases}
\frac{\beta-\alpha}{\beta(1-\alpha)} &\text{if }0\le\alpha<\beta<1\\
\frac{\beta-\alpha}{\alpha(\beta-1)} &\text{if }\beta>\alpha>1
\end{cases}\;.
\ee
Then, $\mu(\eps,\alpha,\beta)=\max\{0,\tmu(\eps,\alpha,\beta)\}$, where
\be
\tmu(\eps,\alpha,\beta)\eqdef\begin{cases}
\frac{\beta}{1-\beta}\left(\theta\log\left(\frac1\eps\right)-h_2(\theta)\right)&\text{if }0\le\alpha<\beta<1\\
\frac{\alpha}{\alpha-1}\left(\theta\log\left(\frac1\eps\right)-h_2(\theta)\right)&\text{if }\beta>\alpha>1\\
0&\text{if }\alpha\geq \beta\geq 0\\
\infty & \text{otherwise}
\end{cases}\;,
\ee
and $h_2(\theta)\eqdef-\theta\log(\theta)-(1-\theta)\log(1-\theta)$ is the binary Shannon entropy.
}

Before proving the theorem, we establish an auxiliary lemma that reduces the optimization in one variable to its two endpoint limits. We will also use it in the proof of Theorem~\ref{thmg1sub}.

\bmyl\label{lem:edge}
Fix $1<\alpha<\beta<\infty$ or $0<\alpha<\beta<1$.
Let $A,B,C,D>0$ and define, for $t\in(0,1]$,
\be\label{eq:phi_def}
\phi(t)\eqdef 
\frac{1}{\beta-1}\log\big(A+Bt^{\beta-1}\big)
-\frac{1}{\alpha-1}\log\big(C+Dt^{\alpha-1}\big).
\ee
Then any critical point $t\in(0,1)$ is a strict local minimum.
Consequently,
\be
\sup_{t\in(0,1]}\phi(t)
=\max\left\{\phi(1),\lim_{t\to0^+}\phi(t)\right\}.
\ee
\emyl

\begin{proof}
Let $g(t)\eqdef A+Bt^{\beta-1}$ and $h(t)\eqdef C+Dt^{\alpha-1}$. 
A direct differentiation gives
\be
(\ln2)\phi'(t)
=
B\frac{t^{\beta-2}}{g(t)}
-
D\frac{t^{\alpha-2}}{h(t)}.
\ee
At a critical point $t\in(0,1)$ we therefore have
\be
s\eqdef \frac{t^{\beta-2}}{g(t)}
=
\frac{D}{B}\frac{t^{\alpha-2}}{h(t)}.
\ee
Using the relation $\phi'(t)=0$ to simplify the second derivative, one obtains
\be
(\ln2)\phi''(t)
=
(\beta-\alpha)B s\Big(\frac{1}{t}-Bs\Big).
\ee
Moreover,
\be
\frac{1}{t}-Bs
=
\frac{1}{t}-B\frac{t^{\beta-2}}{g(t)}
=
\frac{g(t)-Bt^{\beta-1}}{t g(t)}
=
\frac{A}{t g(t)}
>0.
\ee
Since $\beta-\alpha>0$ and $Bs>0$, it follows that $\phi''(t)>0$, and therefore $t$ is a strict local minimum.
\end{proof}

We now prove Theorem~\ref{thmg1}.

\begin{proof}[Proof of Theorem~\ref{thmg1}]
By Lemma~\ref{lemc} and Corollary~\ref{cor001}, it suffices to
consider classical probability vectors with a full-support reference:
\be
\mu(\eps,\alpha,\beta)
=\sup_{d\in\mbb{N}}
\sup_{\substack{\p,\q\in\prob(d)\\q_x>0\ \forall x}}
\left\{D_\beta^\eps(\p\|\q)-D_\alpha(\p\|\q)\right\}.
\ee
Conversely, every such pair gives an admissible pair of commuting
quantum states, so classical examples establish sharpness.
Taking $\p=\q$ gives $\mu(\eps,\alpha,\beta)\ge0$.

\medskip
\noindent\emph{1. The case $\alpha\ge\beta$.}
Smoothing and monotonicity in the R\'enyi order give
\be
D_\beta^\eps(\p\|\q)\le D_\beta(\p\|\q)\le D_\alpha(\p\|\q).
\ee
Hence $\mu(\eps,\alpha,\beta)=0$.

\medskip
\noindent\emph{2. Reduction to binary distributions for
$0<\alpha<\beta<1$ and $1<\alpha<\beta<\infty$.}
Fix $\p,\q\in\prob(d)$ with $\q>0$. If $T(\p,\q)\le\eps$, then
$D_\beta^\eps(\p\|\q)=0$, and the difference is nonpositive.
Otherwise, let $\s=\p^{(\eps)}$, and write
\be
r_x=\frac{p_x}{q_x},
\qquad
s_x=q_x\max\{b,\min\{a,r_x\}\},
\qquad a>1>b>0,
\ee
where the upper and lower transferred masses both equal $\eps$.
By Theorem~\ref{thmg0},
$D_\beta^\eps(\p\|\q)=D_\beta(\s\|\q)$.

We use a \emph{split--merge construction} to obtain a binary
pair whose objective is at least as large as that of the
original pair $\p,\q$. First, we split each coordinate with
$b<r_x<a$ into two coordinates, so that the modified clipped
vector has only the two likelihood ratios $a$ and $b$
relative to the modified reference weights. We then normalize
these weights and merge all coordinates with the same
clipped likelihood ratio into a single coordinate.

To carry out the splitting step, consider a coordinate with
$b<r_x<a$. Choose $\lambda\in(0,1)$ such that
\be
r_x^{\beta-1}
=\lambda a^{\beta-1}+(1-\lambda)b^{\beta-1},
\ee
and replace this coordinate by two coordinates as follows:
\be\label{splitting0}
\begin{aligned}
q_x&\longmapsto
\left(\frac{\lambda p_x}{a},
      \frac{(1-\lambda)p_x}{b}\right),\\
p_x=s_x&\longmapsto
\big(\lambda p_x,(1-\lambda)p_x\big).
\end{aligned}
\ee
The mass of each of $\p$ and $\s$ is clearly preserved,
and the two new likelihood ratios are $a$ and $b$.
Moreover, the contribution to the $\beta$-power sum of $\s$
is unchanged:
\be\label{betasum}
\begin{aligned}
(\lambda p_x)^\beta
 \left(\frac{\lambda p_x}{a}\right)^{1-\beta}
+((1-\lambda)p_x)^\beta
 \left(\frac{(1-\lambda)p_x}{b}\right)^{1-\beta}
&
=p_x\big(\lambda a^{\beta-1}
       +(1-\lambda)b^{\beta-1}\big)\\
&=p_xr_x^{\beta-1}
=s_x^\beta q_x^{1-\beta}.
\end{aligned}
\ee

The contribution of the two new coordinates to the
$\alpha$-power sum of $\p$ is
\be\label{172}
\begin{aligned}
(\lambda p_x)^\alpha
 \left(\frac{\lambda p_x}{a}\right)^{1-\alpha}
+((1-\lambda)p_x)^\alpha
 \left(\frac{(1-\lambda)p_x}{b}\right)^{1-\alpha}
=p_x\big(\lambda a^{\alpha-1}
       +(1-\lambda)b^{\alpha-1}\big).
\end{aligned}
\ee
The function $r\mapsto r^{(\alpha-1)/(\beta-1)}$ is concave
when $1<\alpha<\beta$ and convex when $0<\alpha<\beta<1$.
Applying Jensen's inequality to
$r_x^{\beta-1}=\lambda a^{\beta-1}+(1-\lambda)b^{\beta-1}$
we get the following bounds on the right-hand side of~\eqref{172}:
\be\label{alphacom}
p_x\big(\lambda a^{\alpha-1}
       +(1-\lambda)b^{\alpha-1}\big)
\begin{cases}
\le p_xr_x^{\alpha-1}=p_x^\alpha q_x^{1-\alpha}\quad\text{ if }
&1<\alpha<\beta\\
\ge p_xr_x^{\alpha-1}=p_x^\alpha q_x^{1-\alpha}\quad\text{ if }
&0<\alpha<\beta<1.
\end{cases}
\ee
Thus the $\alpha$-power sum decreases in the first regime
and increases in the second. Since $\alpha-1$ is respectively
positive and negative, its logarithm divided by $\alpha-1$
cannot increase in either regime.

After all replacements, denote the resulting vectors by
$\widetilde{\p},\widetilde{\s},\widetilde{\q}$.
The first two are probability vectors, whereas
$\widetilde{\q}$ need not be normalized. This does not affect
the objective: multiplying the reference vector by any $c>0$
shifts each R\'enyi divergence by $-\log (c)$, independently of
its order, so these shifts cancel in their difference.
We may therefore set
$\widehat{\q}=\widetilde{\q}/\sum_y\widetilde q_y$
without changing the objective; i.e.
\be\label{ineqab}
D_\beta(\widetilde{\s}\|\widehat{\q})
-D_\alpha(\widetilde{\p}\|\widehat{\q})
=
D_\beta(\widetilde{\s}\|\widetilde{\q})
-D_\alpha(\widetilde{\p}\|\widetilde{\q}).
\ee
Now, the preservation of the $\beta$-power sum~\eqref{betasum} gives $D_\beta(\widetilde{\s}\|\widetilde{\q})=D_\beta(\s\|\q)$, and the
$\alpha$-power-sum comparison~\eqref{alphacom} gives $D_\alpha(\widetilde{\p}\|\widetilde{\q})\le D_\alpha(\p\|\q)$. Combining these with~\eqref{ineqab} yields
\ba\label{comparison}
D_\beta(\widetilde{\s}\|\widehat{\q})
-D_\alpha(\widetilde{\p}\|\widehat{\q})
&\ge
D_\beta(\s\|\q)-D_\alpha(\p\|\q)\\
&=D_\beta^\eps(\p\|\q)-D_\alpha(\p\|\q).
\ea
Moreover, $\widetilde{\s}$ remains the flattest
$\eps$-approximation of $\widetilde{\p}$ relative to
$\widehat{\q}$, but we do not need this fact here and omit
its verification. We will verify optimality directly for
the final binary pair below.

After splitting, every coordinate satisfies either
$\widetilde s_y=a\widetilde q_y$ or
$\widetilde s_y=b\widetilde q_y$.
We call the coordinates satisfying the first equality the
upper group, and those satisfying the second the lower group.
We now replace each of the vectors
$\widetilde{\p},\widetilde{\s},\widehat{\q}$
by a binary vector whose entries are the sums over the
upper and lower groups, respectively.

Let $p$ be the sum of the entries of $\widetilde{\s}$ in
the upper group; its mass in the lower group is then $1-p$.
Clipping transfers mass $\eps$ from the upper group to the
lower group, and our splitting operation leaves these
mass differences unchanged. Hence the corresponding masses
of $\widetilde{\p}$ are $p+\eps$ and $1-p-\eps$.
Moreover, the identities defining the groups imply that
their masses in $\widetilde{\q}$ are $p/a$ and $(1-p)/b$.

Put $v=b/a$, and denote the binary probability vectors
obtained by merging $\widetilde{\p}$ and $\widehat{\q}$
by $\r$ and $\t$, respectively. Then
\be\label{t1-binary-family}
\r=(p+\eps,1-p-\eps),
\qquad
\t=\frac{(p/a,(1-p)/b)}{p/a+(1-p)/b}
=\frac{(vp,1-p)}{1-p+vp},
\ee
where $0<p\le1-\eps$ and $0<v<1$.
The vector obtained by merging $\widetilde{\s}$ is $(p,1-p)$.
Since $t_1<p$, transferring mass $\eps$ from the first
coordinate of $\r$ to the second moves $\r$ toward $\t$
without reaching it. This is precisely the binary clipping
construction. Thus the flattest $\eps$-approximation of $\r$
relative to $\t$, denoted by $\r^{(\eps)}$, is
\be
\r^{(\eps)}=(p,1-p).
\ee

For the $\beta$-divergence, the ratio
$\widetilde s_y/\widehat q_y$ is constant within each group,
so merging preserves the divergence. Together with the
identified clipped optimizer $\r^{(\eps)}$, this gives
\be
D_\beta^\eps(\r\|\t)=D_\beta(\r^{(\eps)}\|\t)=D_\beta(\widetilde{\s}\|\widehat{\q}).
\ee
For the $\alpha$-divergence, DPI under the same merging
map gives
\be
D_\alpha(\r\|\t)\le D_\alpha(\widetilde{\p}\|\widehat{\q}).
\ee
Combining these facts gives
\be
\begin{aligned}
F(p,v)
 &\eqdef D_\beta^\eps(\r\|\t)-D_\alpha(\r\|\t)\\
 &\ge D_\beta(\widetilde{\s}\|\widehat{\q})
     -D_\alpha(\widetilde{\p}\|\widehat{\q})\\
&\ge D_\beta^\eps(\p\|\q)-D_\alpha(\p\|\q)\,,
\end{aligned}
\ee
where the second inequality is~\eqref{comparison}.
Thus, for every starting pair $\p,\q$ with
$T(\p,\q)>\eps$, the construction produces binary probability
vectors whose objective is at least as large.
For each fixed $v\in(0,1)$, $F(p,v)$ is continuous at
$p=1-\eps$. Thus its supremum is unchanged if we restrict
to $0<p<1-\eps$.

\medskip
\noindent\emph{3. Evaluation of the binary supremum.}
The preceding reduction shows that every original pair
with $T(\p,\q)>\eps$ is dominated by a binary pair
of the form~\eqref{t1-binary-family}.
Conversely, every such binary pair is admissible in the
original optimization. Together with the value zero
obtained by taking $\p=\q$, this gives
\be
\mu(\eps,\alpha,\beta)
=\max\left\{0,\;
\sup_{\substack{0<p<1-\eps\\0<v<1}}F(p,v)\right\}.
\ee

Substituting the binary vectors into the R\'enyi formulas,
the common term $\log\big((1-p+vp)/v\big)$ cancels, giving
\ba\label{t1-binary-objective}
F(p,v)&=D_\beta(\r^{(\eps)}\|\t)-D_\alpha(\r\|\t)\\
&=\frac1{\beta-1}
\log\bigl(p+(1-p)v^{\beta-1}\bigr)
-\frac1{\alpha-1}
\log\Bigl(p^{1-\alpha}(p+\eps)^\alpha
+(1-p)^{1-\alpha}(1-p-\eps)^\alpha
v^{\alpha-1}\Bigr).
\ea
For fixed $p\in(0,1-\eps)$, all four coefficients in
the two logarithms are positive. Lemma~\ref{lem:edge}
therefore implies that every interior critical point
in $v$ is a strict local minimum. Hence the supremum
over $v$ is the larger of the two endpoint limits.

As $v\to1$, the reference vector $\t$ tends to
$\r^{(\eps)}=(p,1-p)$, and therefore
\be
\lim_{v\to1}F(p,v)
=-D_\alpha(\r\|\r^{(\eps)})\le0.
\ee
To evaluate the other endpoint, define
\be
g_\theta(x)\eqdef
(1-\theta)\log x-\log(x+\eps),
\qquad 0<x<1-\eps,
\ee
where $\theta$ is given in~\eqref{theta}.
For $1<\alpha<\beta$, the powers of $v$ in
\eqref{t1-binary-objective} vanish as $v\to0$.
For $0<\alpha<\beta<1$, factor out $v^{\beta-1}$
and $v^{\alpha-1}$ from the respective logarithms;
the resulting $\log (v)$ terms cancel.
In both cases, using the definition of $\theta$ gives
\be
\lim_{v\to0}F(p,v)=
\begin{cases}
\displaystyle
\frac{\alpha}{\alpha-1}g_\theta(p)
&\text{if }\;\;1<\alpha<\beta,\\[6pt]
\displaystyle
\frac{\beta}{1-\beta}g_\theta(1-p-\eps)
&\text{if }\;\;0<\alpha<\beta<1.
\end{cases}
\ee
Since both $p$ and $1-p-\eps$ range over $(0,1-\eps)$,
the remaining optimization is the same in both regimes.

Differentiating gives
\be
(\ln2)g_\theta'(x)
=\frac{(1-\theta)\eps-\theta x}{x(x+\eps)}.
\ee
If $\eps<\theta$, the derivative changes from positive
to negative at
\be
x_*=\frac{\eps(1-\theta)}{\theta}\in(0,1-\eps),
\ee
so the maximum is
\be
g_\theta(x_*)
=\theta\log\frac1\eps-h_2(\theta).
\ee
If $\eps\ge\theta$, the function is increasing throughout
$(0,1-\eps)$, and its supremum is the boundary value
\be
\sup_{0<x<1-\eps}g_\theta(x)
=(1-\theta)\log(1-\eps)<0.
\ee
In this latter case, the expression appearing in the
theorem is also negative, since
\be
\theta\log\frac1\eps-h_2(\theta)
\le
\theta\log\frac1\theta-h_2(\theta)
=(1-\theta)\log(1-\theta)<0.
\ee
Thus, including the value zero, we obtain precisely
the two finite-order branches of the theorem.

For explicit sharpness when the claimed correction is
positive, we necessarily have $\eps<\theta$.
Choose $p=x_*$ when $1<\alpha<\beta$, and
$p=1-\eps-x_*$ when $0<\alpha<\beta<1$.
In either case $0<p<1-\eps$.
The binary pairs $\r,\t$ in~\eqref{t1-binary-family}
with $v=1/n$, $n\ge2$, then approach the claimed correction.
Each pair has a full-support reference vector.
If the correction is zero, it is attained at $\p=\q$.

\medskip
\noindent\emph{4. The endpoints $\alpha=0$ and $\beta=\infty$.}
First consider $\alpha=0$ and $0<\beta<1$.
For any $\p,\q$, the probability vector
$\p'=(1-\eps)\p+\eps\q$ satisfies
\be
T(\p,\p')=\eps T(\p,\q)\le\eps,
\qquad
\p'\ge\eps\q.
\ee
Consequently,
\be
\sum_x(p_x')^\beta q_x^{1-\beta}
\ge\eps^\beta\sum_xq_x
=\eps^\beta.
\ee
Since $\beta-1<0$ and $D_0(\p\|\q)\ge0$, this gives
\be
D_\beta^\eps(\p\|\q)-D_0(\p\|\q)
\le D_\beta(\p'\|\q)
\le\frac{\beta}{1-\beta}\log\frac1\eps.
\ee

For sharpness, use the binary family
\eqref{t1-binary-family} with
$p=1-\eps-\delta$, where $0<\delta<1-\eps$.
Then
\be
\r=(1-\delta,\delta),
\qquad
\r^{(\eps)}=(1-\eps-\delta,\eps+\delta),
\ee
where the clipping is relative to $\t$.
Both entries of $\r$ are positive, so $D_0(\r\|\t)=0$.
For fixed $\delta$, letting $v\to0$ gives
\be
\lim_{v\to0}
\left[D_\beta^\eps(\r\|\t)-D_0(\r\|\t)\right]
=
\frac{\beta}{1-\beta}\log\frac1{\eps+\delta}.
\ee
Letting $\delta\to0$ proves sharpness of the upper bound,
which is the claimed formula with $\theta=1$.

Now let $1<\alpha<\infty$ and $\beta=\infty$.
The clipped optimizer is independent of the R\'enyi order.
Hence, for every fixed finite-dimensional pair
$\p,\q$ with $\q>0$,
\be
\begin{aligned}
\lim_{\beta\to\infty}D_\beta^\eps(\p\|\q)
&=\lim_{\beta\to\infty}
D_\beta(\p^{(\eps)}\|\q)\\
&=D_{\max}(\p^{(\eps)}\|\q)
=D_{\max}^\eps(\p\|\q).
\end{aligned}
\ee
Taking this pointwise limit in the finite-order upper
bound, with $\theta\to1/\alpha$, gives the claimed upper
bound for $\beta=\infty$.

When the claimed correction is positive, $\eps<1/\alpha$,
so $p=(\alpha-1)\eps$ belongs to $(0,1-\eps)$.
For this choice of $p$, the binary family
\eqref{t1-binary-family} satisfies
\be
\begin{aligned}
&\lim_{v\to0}
\left[D_{\max}^\eps(\r\|\t)-D_\alpha(\r\|\t)\right]\\
&\qquad
=\log p-\frac{\alpha}{\alpha-1}\log(p+\eps)\\
&\qquad
=\frac{\alpha}{\alpha-1}
\left(\frac1\alpha\log\frac1\eps
-h_2\left(\frac1\alpha\right)\right).
\end{aligned}
\ee
This proves sharpness of the positive correction.
The zero case is again attained at $\p=\q$.

\medskip
\noindent\emph{5. The remaining cases.}
It remains to prove unboundedness when
$\alpha<\beta$ and $\alpha\le1\le\beta$.
Fix $0<p<1-\eps$ and set
\be
\p=(p+\eps,1-p-\eps),\qquad
\q_t=(t,1-t),\qquad 0<t<p.
\ee
Then $\p^{(\eps)}=(p,1-p)$.
Writing $L=\log(1/t)$, the defining formulas give, as $t\to0$,
\be
D_\beta^\eps(\p\|\q_t)-D_\alpha(\p\|\q_t)
=
\begin{cases}
L+O(1),&0\le\alpha<1<\beta\le\infty,\\
(1-p-\eps)L+O(1),&\alpha=1<\beta\le\infty,\\
pL+O(1),&0\le\alpha<\beta=1.
\end{cases}
\ee
Every coefficient of $L$ is positive, so the difference
diverges to $+\infty$. This completes all cases.
\end{proof}

\section{Proof of Theorem~\ref{thmg2}}

\noindent\textbf{Theorem 2.} {\it Let $\eps\in(0,1)$, $\alpha,\beta\in[0,\infty]$, and
\be
\nu(\eps,\alpha,\beta)\eqdef\sup_A\sup_{\rho,\sigma\in\md(A)}
\left\{D_\alpha^{\mbb{M}}(\rho\|\sigma)-\D_\beta^\eps(\rho\|\sigma)\right\}\;.
\ee
Then
\be
\nu(\eps,\alpha,\beta)=
\begin{cases}
\left(\frac1{\beta-1}+\frac1{1-\alpha}\right)\log\frac1{1-\eps}
& \text{if }\beta>1>\alpha\\[6pt]
\infty & \text{otherwise}\;.
\end{cases}
\ee}

\begin{proof}
By Lemma~\ref{lemc}, it suffices to bound the classical difference. For commuting inputs, dephasing shows that smoothing any quantum extension of $D_\beta$ gives the classical smoothed divergence. Thus classical examples also establish sharpness for every quantum extension in the theorem.

\noindent\emph{Finite branch: $0\le\alpha<1<\beta\le\infty$.}
Fix $\p,\q$ and $\p'\in\mb^\eps(\p)$. Set $r_x\eqdef\min\{p_x,p'_x\}$, $r\eqdef\sum_xr_x$, and $\s\eqdef\r/r$. Observe that
\be
r=\sum_x\min\{p_x,p'_x\}=1-\frac12\|\p-\p'\|_1\geq1-\eps\,.
\ee
 Monotonicity in each component of the first argument, and monotonicity in the R\'enyi order give
\ba\label{overlap-bound}
D_\alpha(\p\|\q)
&\le D_\alpha(\r\|\q)\\
&=D_\alpha(\s\|\q)+\frac{\alpha}{\alpha-1}\log (r)\\
&\le D_\beta(\s\|\q)+\frac{\alpha}{\alpha-1}\log (r)\\
&=D_\beta(\r\|\q)
+\left(\frac1{1-\alpha}+\frac1{\beta-1}\right)\log\frac1r\\
&\le D_\beta(\p'\|\q)
+\left(\frac1{1-\alpha}+\frac1{\beta-1}\right)\log\frac1{1-\eps}\,
\ea
where in the last inequality we used the monotonicity of $D_\beta$ (with $\beta>1$) in each component of the first argument.
At $\alpha=0$, the first inequality follows from inclusion of supports and the homogeneity term is zero. At $\beta=\infty$, use $D_{\max}(r\s\|\q)=D_{\max}(\s\|\q)+\log (r)$. Minimizing over $\p'$ proves the claimed upper bound on $\nu$.

For sharpness, let $k\to\infty$, $\ell=k^2$, $a=(1-\eps)/k$, and $b=\eps/\ell$. In dimension $d=k+\ell$, set
\be\label{lower-sharp-sequence}
\p^{(k)}=(a+\eps,\underbrace{a,\ldots,a}_{k-1},
\underbrace{0,\ldots,0}_{\ell}),\qquad
\upb^{(\eps,k)}=(\underbrace{a,\ldots,a}_{k},
\underbrace{b,\ldots,b}_{\ell}).
\ee
For large $k$, $a>b$ and the second vector is exactly the $\eps$-clipped vector of the first. Moreover,
$T(\p^{(k)},\u_d)=\ell/d>\eps$.
For $0<\alpha<1$ and $1<\beta<\infty$,
\ba
H_\alpha(\p^{(k)})&=\log (k)+\frac{\alpha}{1-\alpha}\log(1-\eps)+o(1),\\
H_\beta(\upb^{(\eps,k)})&=\log (k)-\frac{\beta}{\beta-1}\log(1-\eps)+o(1).
\ea
Thus $H_\beta(\upb^{(\eps,k)})-H_\alpha(\p^{(k)})$ approaches the claimed value. For the endpoints,
$H_0(\p^{(k)})=\log (k)$ and
$H_\infty(\upb^{(\eps,k)})=\log (k)-\log(1-\eps)$, giving the same formula directly.

\medskip
\noindent\emph{Infinite branches.}
If $\alpha\ge1$, take
\be
\p^{(d)}=\left(\eps,\frac{1-\eps}{d-1},\ldots,\frac{1-\eps}{d-1}\right).
\ee
For large $d$, $T(\p^{(d)},\u_d)=\eps-1/d<\eps$, so $D_\beta^\eps(\p^{(d)}\|\u_d)=0$ for every $\beta$. Meanwhile \be D_1(\p^{(d)}\|\u_d)=\eps\log (d)-h_2(\eps)+o(1)\to\infty\,,\ee and $D_\alpha\ge D_1$.
For every remaining case, $\beta\le1$. Take $\p=\e_1$ in dimension $d$; its clipped vector is
$(1-\eps,\eps/(d-1),\ldots,\eps/(d-1))$ for large $d$.
Since $D_\alpha(\e_1\|\u_d)=\log (d)$ for every $\alpha$, the difference equals the entropy of this clipped vector. It is at least its Shannon entropy,
$h_2(\eps)+\eps\log(d-1)\to\infty$, for every $0\le\beta\le1$.
These admissible families cover all infinite branches, including orders zero and one.
\end{proof}

The case $\beta=\infty$ connects Theorem~\ref{thmg2} with
the strong converse results for smoothed max-relative
entropy studied by Salzmann and Datta~\cite{SD2022} and
Li and Yao~\cite{LY2024}. Our theorem uses trace-distance
smoothing over normalized states, whereas Li and Yao's
exact exponent concerns purified-distance smoothing over
subnormalized states. Moreover, our correction is optimal
uniformly over all states and dimensions, while a strong
converse exponent describes tensor powers of a fixed pair.
The sharpness examples establishing universal optimality
need not have this tensor-power structure. Thus
Theorem~\ref{thmg2} yields an asymptotic lower bound, but
its universal sharpness alone does not establish an exact
strong converse exponent.

\section{Proof of Theorem~\ref{thmg3}}

\noindent\textbf{Theorem 3.} {\it Let $\eps\in(0,1)$, $\alpha\in[0,\infty]$, and
\be
\mu_H(\eps,\alpha)\eqdef\sup_A\sup_{\rho,\sigma\in\md(A)}\left\{D_H^{\eps}(\rho\|\sigma)- \D_\alpha(\rho\|\sigma)\right\}\;,
\ee
where $\D_\alpha$ is any quantum extension of the R\'enyi relative entropy of order $\alpha$. Then:
\be
\mu_H(\eps,\alpha)= \begin{cases}\frac\alpha{\alpha-1}\log\left(\frac1{1-\eps}\right) &\text{if }\alpha>1\\
\infty &\text{otherwise.}
\end{cases}
\ee}

\begin{proof}
For $\alpha>1$, this universal upper bound is well known, so it remains only to prove achievability (and hence optimality).
Consider the probability vector
\be
\p\eqdef\left(1-\eps,\frac\eps{d-1},\ldots,\frac\eps{d-1}\right)^T\;.
\ee
Since $\eps<1$, for all sufficiently large $d$ we have $\p=\p^\da$. Taking $\q=\u$ and using~\eqref{formula}, we note that in this case $\ell=0$, so that $a_\ell=b_\ell=0$. Hence
\be
D_H^\eps(\p\|\u)=\log(d)\;.
\ee
On the other hand,
\be
D_\alpha(\p\|\u)
=
\log(d)+\frac{\log\left((1-\eps)^\alpha+(d-1)^{1-\alpha}\eps^\alpha\right)}{\alpha-1}.
\ee
Therefore,
\ba
D_H^\eps(\p\|\u)-D_\alpha(\p\|\u)
&=
-\frac{\log\left((1-\eps)^\alpha+(d-1)^{1-\alpha}\eps^\alpha\right)}{\alpha-1}\\
&\xrightarrow[d\to\infty]{}
-\frac1{\alpha-1}\log\left((1-\eps)^\alpha\right)\\
&=
\frac\alpha{\alpha-1}\log\frac1{1-\eps}\;.
\ea
Thus the finite value is attained as a supremum. At $\alpha=\infty$, the same family gives the difference $-\log(1-\eps)$ exactly for large $d$.
For $0\le\alpha\le1$, the difference for this family is $H_\alpha(\p)$, which is at least
$H_1(\p)=h_2(\eps)+\eps\log(d-1)\to\infty$.
This proves every branch directly.
\end{proof}

\section{Proof of Theorem~\ref{thmg4}}

\noindent\textbf{Theorem 4.} {\it Let $\eps\in(0,1)$, $\alpha\in[0,\infty]$, and
\be\label{2000}
\nu_H(\eps,\alpha)\eqdef\sup_A\sup_{\rho,\sigma\in\md(A)}\left\{D_\alpha^{\mbb{M}}(\rho\|\sigma)-D^{\eps}_H(\rho\|\sigma)\right\}.
\ee
Then:
\be
\nu_H(\eps,\alpha)=\begin{cases}
-\log\frac1{1-\eps} & \text{ if }\alpha\in[0,\eps]\\
\frac{\alpha}{1-\alpha}\log\left(\frac\alpha\eps\right)-\log\frac1{1-\alpha} & \text{ if }\alpha\in(\eps,1)\\
\infty  & \text{ if }\alpha\in[1,\infty]\;.
\end{cases}
\ee}

\begin{proof}
By Lemma~\ref{lemc} and Corollary~\ref{cor001}, the optimal constant is the supremum of
$D_\alpha(\p\|\u)-D_H^\eps(\p\|\u)$ over finite probability vectors.

\noindent\emph{The branch $0\le\alpha\le\eps$.}
Apply the known bound~\eqref{oub} at order $\eps$ and use monotonicity in the order:
\be
D_H^\eps(\p\|\u)
\ge D_\eps(\p\|\u)-\log(1-\eps)
\ge D_\alpha(\p\|\u)-\log(1-\eps).
\ee
Equality in the resulting difference is attained at $\p=\u$, where
$D_H^\eps(\u\|\u)=-\log(1-\eps)$ and $D_\alpha(\u\|\u)=0$.
This includes $\alpha=0$ and $\alpha=\eps$ without an interchange of an infinite limit and a supremum.

\noindent\emph{The branch $\eps<\alpha<1$.}
The upper bound on the difference is~\eqref{oub}. For sharpness, let
\be\label{hyp-lower-sharp}
\p^{(m)}=\left(1-\frac\eps\alpha,
\underbrace{\frac\eps{m\alpha},\ldots,\frac\eps{m\alpha}}_{m}\right),
\qquad \q=\u_{m+1}.
\ee
For sufficiently large $m$, the first atom has the largest likelihood ratio. An optimal test accepts it and accepts a total of $m(1-\alpha)$ tail coordinates, allowing a fractional acceptance probability for the last coordinate. Its success probability is exactly $1-\eps$, and its type-II error is
$[1+m(1-\alpha)]/(m+1)$. Consequently,
\ba
&D_\alpha(\p^{(m)}\|\u_{m+1})-D_H^\eps(\p^{(m)}\|\u_{m+1})\\
&=\log[1+m(1-\alpha)]
-\frac1{1-\alpha}\log\left[\left(1-\frac\eps\alpha\right)^\alpha
+m^{1-\alpha}\left(\frac\eps\alpha\right)^\alpha\right]\\
&\longrightarrow\log(1-\alpha)+\frac\alpha{1-\alpha}\log\frac\alpha\eps.
\ea
All block sizes are integers, and the fractional test is allowed by the definition of $D_H^\eps$; no relaxation of the distribution is used.

\noindent\emph{The branch $\alpha\ge1$.}
Fix $0<\eta<1-\eps$ and set
\be
\p^{(m)}=\Big(1-\eps-\eta,
\underbrace{\frac{\eps+\eta}{m},\ldots,\frac{\eps+\eta}{m}}_{m}\Big).
\ee
For large $m$, the optimal test accepts the first atom and a fraction $\eta/(\eps+\eta)$ of the tail, so
\be
D_H^\eps(\p^{(m)}\|\u_{m+1})
=-\log\frac{1+m\eta/(\eps+\eta)}{m+1}
\longrightarrow\log\frac{\eps+\eta}{\eta}.
\ee
In contrast,
$D_1(\p^{(m)}\|\u_{m+1})=(1-\eps-\eta)\log (m)-h_2(\eps+\eta)+o(1)\to\infty$.
Monotonicity $D_\alpha\ge D_1$ proves the infinite branch for every $\alpha\in[1,\infty]$.
\end{proof}

\section{Proof of Theorem~\ref{thmg1sub}}

\noindent\textbf{Theorem 5.} {\it Let $\eps\in(0,1)$ and $\beta>\alpha>1$. Then:
\be\label{mainforsuba}
\mu_\sub(\eps,\alpha,\beta)=\begin{cases}
\frac{\alpha}{\alpha-1}\left(\theta\log\left(\frac1\eps\right)-h_2(\theta)\right)&\text{if }\;\;\eps\le\theta\\
\frac{\beta}{\beta-1}\log(1-\eps)&\text{if }\;\;\eps>\theta
\end{cases}
\ee
Moreover, for $0<\alpha<\beta<1$ we have $\mu_\sub(\eps,\alpha,\beta)=\mu(\eps,\alpha,\beta)$.}

\begin{proof}
By Lemma~\ref{lemc} and the full-support approximation used
in Corollary~\ref{cor001},
\be\label{msub}
\mu_\sub(\eps,\alpha,\beta)
=
\sup_{d\in\mbb{N}}
\sup_{\substack{\p,\q\in\prob(d)\\ \q>0}}
\Big\{
D_\beta^{\eps,\le}(\p\|\q)-D_\alpha(\p\|\q)
\Big\}.
\ee
We first consider $1<\alpha<\beta<\infty$.

\medskip
\noindent\emph{1. The subnormalized optimizer.}
Fix $\p,\q\in\prob(d)$ with $\q>0$.
For a subnormalized vector $\r$, the generalized trace
distance from $\p$ is
\be
\|\p-\r\|_+=\sum_x(p_x-r_x)_+.
\ee
Replacing any feasible $\r$ by
$r'_x=\min\{r_x,p_x\}$ cannot increase
$D_\beta(\r\|\q)$, since $\beta>1$. Moreover,
\be
\sum_xr'_x
=
1-\sum_x(p_x-r_x)_+
\ge1-\eps.
\ee
Scaling $\r'$ by $(1-\eps)/\sum_xr'_x$ therefore gives
a vector below $\p$ with mass exactly $1-\eps$,
without increasing the divergence. Its generalized
trace distance from $\p$ is exactly $\eps$, so it
remains feasible. Thus we may restrict the minimization to
\be\label{r209}
0\le r_x\le p_x,
\qquad
\sum_xr_x=1-\eps.
\ee

Let $a>0$ be the unique solution of
\be
\sum_x(p_x-aq_x)_+=\eps,
\ee
and define the capped vector $\s$ by
\be\label{t5-cap}
s_x=\min\{p_x,aq_x\}.
\ee
By definition, $\sum_xs_x=1-\eps$.
The vector $\s$ is optimal. Since $\beta>1$, it suffices
to show that $\s$ minimizes the $\beta$-power sum.
For every $\r$ satisfying~\eqref{r209}, we apply the
convexity inequality
\be
z^\beta-z_0^\beta
\ge \beta z_0^{\beta-1}(z-z_0)
\ee
with $z=r_x/q_x$ and $z_0=s_x/q_x$.
Multiplying by $q_x$ and summing over all coordinates
gives the first inequality below:
\be
\begin{aligned}
\sum_x r_x^\beta q_x^{1-\beta}
-\sum_x s_x^\beta q_x^{1-\beta}
&\ge
\beta\sum_x
\left(\frac{s_x}{q_x}\right)^{\beta-1}(r_x-s_x)\\
&\ge
\beta a^{\beta-1}\sum_x(r_x-s_x)
=0.
\end{aligned}
\ee
For the second inequality, consider each coordinate.
If $s_x=aq_x$, its coefficient is already $a^{\beta-1}$.
Otherwise, $s_x=p_x<aq_x$, so
$r_x-s_x\le0$ and
$(s_x/q_x)^{\beta-1}<a^{\beta-1}$.
Multiplying the latter inequality by $r_x-s_x\le0$
reverses its direction, giving the required bound.
Finally, $\sum_x(r_x-s_x)=0$ because both vectors
have mass $1-\eps$.
Consequently,
\be
D_\beta^{\eps,\le}(\p\|\q)=D_\beta(\s\|\q).
\ee

Since the optimizer can be chosen below $\p$, coordinates
with $p_x=0$ may be removed. Renormalizing $\q$ on the
remaining support shifts both divergences by the same
constant, leaving their difference unchanged.
We may therefore assume $\p>0$ as well.
Moreover,
if $\s$ is proportional to $\q$, then
$\s=(1-\eps)\q$, and
\be
D_\beta^{\eps,\le}(\p\|\q)-D_\alpha(\p\|\q)
=
\frac{\beta}{\beta-1}\log(1-\eps)
-D_\alpha(\p\|\q)
\le
\frac{\beta}{\beta-1}\log(1-\eps).
\ee
This value is attained when $\p=\q$.
It remains to consider the case in which $\s$ is not
proportional to $\q$.

\medskip
\noindent\emph{2. Reduction to binary distributions.}
Put
\be
r_x=\frac{p_x}{q_x},
\qquad
b=\min_x r_x.
\ee
Since $\p,\q>0$, we have $b>0$.
Moreover, $b<a$; otherwise $s_x=aq_x$ for every $x$,
contrary to the assumption that $\s$ is not proportional
to $\q$.

Apply the coordinate-splitting construction~\eqref{splitting0} from the proof
of Theorem~\ref{thmg1} to every coordinate with $b<r_x<a$,
where $p_x=s_x$.
Denote the resulting vectors by
$\widetilde{\p},\widetilde{\s},\widetilde{\q}$, and set
$\widehat{\q}=\widetilde{\q}/\sum_y\widetilde q_y$.
The same power-sum comparisons apply: splitting preserves
the $\beta$-power sum of $\s$ and cannot increase the
$\alpha$-power sum of $\p$.
These comparisons require only $0<b<a$ and
$1<\alpha<\beta$, and do not require $\s$ to be normalized.
Likewise, normalizing the reference weights shifts both
divergences by the same constant. Thus the difference
cannot decrease.

After splitting, every coordinate satisfies either
$\widetilde s_y=a\widetilde q_y$ or
$\widetilde s_y=b\widetilde q_y$.
Call these the upper and lower groups, respectively.
Let $p$ be the mass of $\widetilde{\s}$ in the upper group;
its lower-group mass is then $1-\eps-p$.
All the removed mass $\eps$ comes from the upper group,
and splitting coordinates with $p_x=s_x$ leaves these
mass differences unchanged. Hence the corresponding
masses of $\widetilde{\p}$ are $p+\eps$ and $1-\eps-p$.
The reference masses in $\widetilde{\q}$ are
$p/a$ and $(1-\eps-p)/b$.

Merge the coordinates within each group and put $v=b/a$.
The resulting probability vectors and capped vector are
\be\label{t5-binary-family}
\begin{aligned}
\r&=(p+\eps,1-\eps-p),\\
\t&=\frac{(vp,1-\eps-p)}{vp+1-\eps-p},\\
\r^{(\eps,\le)}&=(p,1-\eps-p),
\end{aligned}
\ee
where $0<p<1-\eps$ and $0<v<1$.
Here $\r^{(\eps,\le)}$ is the capped optimizer for $\r$
relative to $\t$. Indeed,
\be
\frac{1-\eps-p}{t_2}
=
v\,\frac{p}{t_1}
<
\frac{p}{t_1}.
\ee
Thus capping $\r$ at the likelihood ratio $p/t_1$
removes mass $\eps$ from its first coordinate and
leaves its second coordinate unchanged.

As in the proof of Theorem~\ref{thmg1}, merging preserves
the $\beta$-divergence because
$\widetilde s_y/\widehat q_y$ is constant within each
group, while DPI can only decrease the
$\alpha$-divergence. Consequently,
\be
\begin{aligned}
F_\sub(p,v)
&\eqdef
D_\beta^{\eps,\le}(\r\|\t)-D_\alpha(\r\|\t)\\
&\ge
D_\beta(\widetilde{\s}\|\widehat{\q})
-D_\alpha(\widetilde{\p}\|\widehat{\q})\\
&\ge
D_\beta^{\eps,\le}(\p\|\q)-D_\alpha(\p\|\q).
\end{aligned}
\ee
Thus every starting pair with a nonproportional capped
vector is dominated by a binary pair.

\medskip
\noindent\emph{3. Evaluation and sharpness.}
For the binary vectors in~\eqref{t5-binary-family},
the common likelihood-ratio scale cancels, giving
\be\label{t5-binary-objective}
\begin{aligned}
F_\sub(p,v)
&=\frac1{\beta-1}
\log\bigl(p+(1-\eps-p)v^{\beta-1}\bigr)\\
&\quad-\frac1{\alpha-1}
\log\Bigl(
p^{1-\alpha}(p+\eps)^\alpha
+(1-\eps-p)v^{\alpha-1}
\Bigr).
\end{aligned}
\ee
For fixed $p\in(0,1-\eps)$, Lemma~\ref{lem:edge}
shows that the supremum over $v$ is determined by
the two endpoint limits.
At $v=1$, we have
$\r^{(\eps,\le)}=(1-\eps)\t$, so
\be
F_\sub(p,1)
=
\frac{\beta}{\beta-1}\log(1-\eps)
-D_\alpha(\r\|\t)
\le
\frac{\beta}{\beta-1}\log(1-\eps).
\ee
At the other endpoint,
\be
\begin{aligned}
\lim_{v\to0}F_\sub(p,v)
&=
\frac{\beta}{\beta-1}\log p
-\frac{\alpha}{\alpha-1}\log(p+\eps)\\
&=
\frac{\alpha}{\alpha-1}g_\theta(p),
\end{aligned}
\ee
where
$g_\theta(p)=(1-\theta)\log p-\log(p+\eps)$
is the function already optimized in the proof of
Theorem~\ref{thmg1}.

The upper bound at $v=1$, which also covers the
proportional case, equals
\be
\lim_{p\to1-\eps}
\frac{\alpha}{\alpha-1}g_\theta(p)
=
\frac{\beta}{\beta-1}\log(1-\eps).
\ee
Since every pair in~\eqref{t5-binary-family} is admissible,
the reduction and these endpoint limits imply
\be
\mu_\sub(\eps,\alpha,\beta)
=
\frac{\alpha}{\alpha-1}
\sup_{0<p<1-\eps}g_\theta(p).
\ee
The scalar maximization from the proof of
Theorem~\ref{thmg1} gives
\be
\sup_{0<p<1-\eps}g_\theta(p)
=
\begin{cases}
\theta\log(1/\eps)-h_2(\theta),
&\eps\le\theta,\\[2pt]
(1-\theta)\log(1-\eps),
&\eps>\theta.
\end{cases}
\ee
This proves~\eqref{mainforsub} for finite $\beta$.

For explicit sharpness when $\eps<\theta$, take
$p=\eps(1-\theta)/\theta$ in~\eqref{t5-binary-family}
and $v=1/n$, $n\ge2$.
These binary pairs approach the first branch of
\eqref{mainforsub}, whether it is positive or negative.
When $\eps\ge\theta$, the choice $\p=\q$ attains
$\frac{\beta}{\beta-1}\log(1-\eps)$.

\medskip
\noindent\emph{4. The endpoint $\beta=\infty$.}
For every fixed pair with $\q>0$, the capped vector
in~\eqref{t5-cap} is independent of $\beta$.
Taking $\beta\to\infty$ in its minimizing inequality
shows that it also minimizes $D_{\max}$.
Consequently,
\be
D_\beta^{\eps,\le}(\p\|\q)
=
D_\beta(\s\|\q)
\longrightarrow
D_{\max}(\s\|\q)
=
D_{\max}^{\eps,\le}(\p\|\q).
\ee
Taking this pointwise limit in the finite-order bound
proves the upper bound with $\theta=1/\alpha$ and
$\beta/(\beta-1)=1$.

For sharpness when $\eps<1/\alpha$, use
\eqref{t5-binary-family} with $p=(\alpha-1)\eps$.
Then
\be
\lim_{v\to0}
\Bigl[
D_{\max}^{\eps,\le}(\r\|\t)
-D_\alpha(\r\|\t)
\Bigr]
=
\log (p)-\frac{\alpha}{\alpha-1}\log(p+\eps),
\ee
which equals the first branch of~\eqref{mainforsub}
at $\theta=1/\alpha$.
For $\eps\ge1/\alpha$, the choice $\p=\q$ attains
$\log(1-\eps)$.

\medskip
\noindent\emph{5. The case $0<\alpha<\beta<1$.}
Let $\r$ be any subnormalized vector feasible for
smoothing $\p$. Complete it to a probability vector
$\s\ge\r$ by adding mass $1-\sum_xr_x$.
Then
\be
\frac12\|\p-\s\|_1
=
\sum_x(p_x-s_x)_+
\le
\sum_x(p_x-r_x)_+
\le\eps.
\ee
For $0<\beta<1$, increasing the first argument
componentwise cannot increase $D_\beta$, since its
power sum increases and $1/(\beta-1)<0$. Hence
\be
D_\beta(\s\|\q)\le D_\beta(\r\|\q).
\ee
Every subnormalized candidate therefore admits a
normalized candidate with no larger divergence.
The reverse inequality follows from inclusion of
the normalized smoothing set, so
\be
D_\beta^{\eps,\le}(\p\|\q)
=
D_\beta^\eps(\p\|\q).
\ee
Subtracting $D_\alpha(\p\|\q)$ and taking the supremum
gives
\be
\mu_\sub(\eps,\alpha,\beta)
=
\mu(\eps,\alpha,\beta).
\ee
Theorem~\ref{thmg1} supplies the claimed value and
its sharpness.
\end{proof}

\section{Proof of Theorem~\ref{thmg0}}\label{Sec9}

\medskip
\noindent\textbf{Theorem~\ref{thmg0} (structural formulation).}
{\it
The vector $\p^{(\eps)}$ defined in Sec.~\ref{sec:arm} is the flattest
$\eps$-approximation of $\p$ relative to $\q$.
Consequently, every classical divergence $\D$ satisfies
\be
\D^\eps(\p\|\q)
=
\D\big(\p^{(\eps)}\big\|\q\big).
\ee
}

\begin{proof}
If $\eps=0$, the statement is immediate.
If $\frac12\|\p-\q\|_1\le\eps$, then
$\p^{(\eps)}=\q\in\mb^\eps(\p)$.
The column-stochastic matrix whose columns all equal $\q$
maps both $\p'$ and $\q$ to $\q$, for every
$\p'\in\prob(d)$. Hence
\be
(\p',\q)\succ(\q,\q)
=(\p^{(\eps)},\q),
\ee
and the conclusion follows from DPI.

We therefore assume
\be
0<\eps<\frac12\|\p-\q\|_1.
\ee
The construction of $\p^{(\eps)}$ removes total mass $\eps$
from the upper part, including the coordinates where $q_x=0$,
and adds total mass $\eps$ to the lower part.
Since $a>1>b$, these parts are disjoint. Consequently,
\be
\p^{(\eps)}\in\prob(d),
\qquad
\frac12\|\p-\p^{(\eps)}\|_1=\eps.
\ee

For arbitrary $\q$, including reference vectors with zero
components, we use the testing-region characterization of
relative majorization~\cite[Theorem~4.10]{Gour2025}.
For normalized probability vectors, it suffices to show
\be\label{testing-criterion}
\sum_{x\in[d]}(p_x'-\lambda q_x)_+
\ge
\sum_{x\in[d]}(p_x^{(\eps)}-\lambda q_x)_+
\qquad\forall\,\lambda\ge0.
\ee

The clipping construction gives the identity
\be\label{clipped-testing-curve}
\sum_{x\in[d]}(p_x^{(\eps)}-\lambda q_x)_+
=
\max\left\{
\sum_{x\in[d]}(p_x-\lambda q_x)_+-\eps,\,
1-\lambda,\,
0
\right\}.
\ee
Indeed, for $0\le\lambda\le b$, the left-hand side is
$1-\lambda$, since $\p^{(\eps)}\ge b\q$.
Moreover, $\sum_x(\lambda q_x-p_x)_+\le\eps$, so the first
term in the maximum is at most $1-\lambda$.
For $b\le\lambda\le a$, removing the upper mass reduces
the positive part by $\eps$, while the mass added below
$b\q$ contributes nothing. The left-hand side is therefore
$\sum_x(p_x-\lambda q_x)_+-\eps$; as a positive-part sum
of a normalized pair, it is at least both $1-\lambda$ and zero.
For finite $a$ and $\lambda\ge a$, the left-hand side vanishes
because $\p^{(\eps)}\le a\q$, while
$\sum_x(p_x-\lambda q_x)_+\le\eps$ and $1-\lambda<0$.
When $a=\infty$, the upper mass is removed entirely outside
$\supp(\q)$, so the middle argument applies to every
$\lambda\ge b$.

For every $\p'\in\mb^\eps(\p)$, the variational
characterization of the positive part implies
\be
\begin{aligned}
\sum_{x\in[d]}(p_x'-\lambda q_x)_+
&\ge
\sum_{x\in[d]}(p_x-\lambda q_x)_+
-\frac12\|\p'-\p\|_1\\
&\ge
\sum_{x\in[d]}(p_x-\lambda q_x)_+-\eps.
\end{aligned}
\ee
Normalization also gives
\be
\sum_{x\in[d]}(p_x'-\lambda q_x)_+
\ge\max\{1-\lambda,0\}.
\ee
Together with~\eqref{clipped-testing-curve}, these inequalities
prove~\eqref{testing-criterion}. Therefore,
\be\label{ppr}
(\p',\q)\succ(\p^{(\eps)},\q)
\qquad\forall\,\p'\in\mb^\eps(\p).
\ee
Finally, DPI implies
\be
\D(\p'\|\q)\ge\D(\p^{(\eps)}\|\q)
\qquad\forall\,\p'\in\mb^\eps(\p).
\ee
Since $\p^{(\eps)}\in\mb^\eps(\p)$, we conclude that
\be
\D^\eps(\p\|\q)
=\min_{\p'\in\mb^\eps(\p)}\D(\p'\|\q)
=\D(\p^{(\eps)}\|\q).
\ee
\end{proof}

\section*{Acknowledgment}
This research was supported by the Israel Science Foundation under Grant No.~1192/24.

\section*{Conflict of Interest}
The author declares that there is no conflict of interest.

\section*{Data Availability}
Data sharing is not applicable to this article as no datasets were generated or analyzed during the current study.

% Bibliography embedded so this source compiles independently of QRT.bib.

\end{document}